\documentclass{jfm}
\usepackage{graphicx}
\usepackage{epstopdf,epsfig}
\usepackage{float} 
\usepackage{newtxtext}
\usepackage[varvw]{newtxmath}
\usepackage[square,sort,comma,numbers]{natbib}
\usepackage{hyperref}
\usepackage{caption}
\usepackage{subcaption}
\usepackage{array}
\usepackage{amsmath}
\hypersetup{
    colorlinks = true,
    urlcolor   = blue,
    citecolor  = black,
}

\newcommand{\RomanNumeralCaps}[1]
\nolinenumbers

\newcommand{\uu }{\boldsymbol{u}}
\newcommand{\nab }{\boldsymbol{\nabla}}
\newcommand{\br}{\boldsymbol{r}}

\newcommand{\Tp}{T}
\newcommand{\rp}{\br}

\title{Large-scale convective flow sustained by thermally active Lagrangian tracers}

\author{Lokahith Agasthya\aff{1} \aff{2} \aff{3}
  \corresp{\email{lnagasthya@gmail.com}},
  Andreas Bartel\aff{2}, 
  Luca Biferale\aff{1}, 
  Matthias Ehrhardt\aff{2}
 \and Federico Toschi\aff{4}}

\affiliation{\aff{1}Department of Physics, University of Rome ``Tor Vergata" and INFN, Via della Ricerca Scientifica~1, 00133, Rome RM, Italy
\aff{2}Angewandte Mathematik und Numerische Analysis, Bergische Universit\"{a}t Wuppertal, Gaußstrasse~20 D-42119 Wuppertal, Germany
\aff{3}Computation-based Science and Technology Research Center, The Cyprus Institute, 20~Kavafi Str., Nicosia 2121, Cyprus
\aff{4}Department of Applied Physics, Eindhoven University of Technology, The Netherlands
}

\begin{document}
\maketitle

\begin{abstract}
Non-isothermal particles suspended in a fluid lead to complex interactions -- the particles respond to changes in the fluid flow, which in turn is modified by their temperature anomaly. Here, we perform a novel proof-of-concept numerical study based on tracer particles that are thermally coupled to the fluid. We imagine that particles can adjust their internal temperature reacting to some local fluid properties and follow simple, hard-wired active control protocols. We study the case where instabilities are induced by switching the particle temperature from hot to cold depending on whether it is ascending or descending in the flow. A macroscopic transition from a stable to unstable convective flow is achieved, depending on the number of active particles and their excess negative/positive temperature. The stable state is characterized by a flow with low turbulent kinetic energy, strongly stable temperature gradient, and no large-scale features. The convective state is characterized by higher turbulent kinetic energy, self-sustaining large-scale convection, and weakly stable temperature gradients. The particles individually promote the formation of stable temperature gradients, while their aggregated effect induces large-scale convection. When the Lagrangian temperature scale is small, a weakly convective laminar system forms. The Lagrangian approach is also compared to a uniform Eulerian bulk heating with the same mean injection profile and no such transition is observed. Our empirical approach shows that thermal convection can be controlled by pure Lagrangian forcing and opens the way for other data-driven particle-based protocols to enhance or deplete large-scale motion in thermal flows.
\end{abstract}



\section{Introduction}
Thermally driven flows play an important role in both nature and industry. 
They are notoriously hard to predict and  control. 
In the presence of gravity, temperature fluctuations cause density fluctuations, which in turn drive convective motions through buoyancy in the atmosphere \citep{salesky2018buoyancy, markowski2007overview}, in oceans \citep{marshall1999open}, and especially in idealized  systems such as Rayleigh-B\'enard convection \citep{ahlers2009heat, lohse2010small}, and horizontal convection \citep{gayen2014stability}. 

It is well known that the two-way interactions between particles suspended in a fluid and the fluid phase itself are complex and highly nonlinear. 
They exhibit behaviour such as preferential concentration due to ejection from vortical regions \citep{squires1991preferential, cencini2006dynamics} and modification of turbulence \citep{yang2005two}. 
The dynamics of particles suspended in turbulence plays an important role in several natural as well as industrial processes, for example in the dispersal of pollutants \citep{fernando2010flow}, clouds \citep{falkovich2002acceleration, mazin1999cloud}, planet formation \citep{bec2014gravity}, combustion of jet sprays \citep{irannejad2015large}.

When suspended particles are thermally coupled to the fluid and are  non-isothermal, the particles cause local temperature fluctuations in the fluid, which in turn can further modify a turbulent flow, either purely by thermal action or also in conjunction with the momentum-coupling \citep{carbone2019multiscale} while momentum coupling alone can also alter the heat-transfer dynamics of a thermal flow \citep{elperin1996turbulent}. 
Modification of specific thermal flows due to suspended, thermal particles has also been studied, for example in the Rayleigh-B\'enard convection \citep{modifiedrayleigh-benard}, where heavy particles with fixed initial temperatures are introduced into a Rayleigh-B\'enard convection system. 
In this case, the particles are found to enhance vertical heat transfer, an effect that is most pronounced when the particle concentration is greatest due to turbulence (preferential concentration), while attenuating turbulent kinetic energy due to momentum-coupling. 
Furthermore, the feasibility of achieving control of Rayleigh-B\'enard convection solely by applying small temperature or velocity fluctuations has been studied \citep{tang1994stabilization}. 
Here, deviations from the stable profiles near the thermal boundaries are detected and compensated, leading to stable Rayleigh-B\'enard flows well above the critical Rayleigh number and also the possibility of control of flow patterns is given. 
Increasing the critical Rayleigh number and delaying the onset of convection can further be improved by applying reinforcement learning techniques to apply the temperature fluctuations near the boundary, as shown by \citep{Luca2020JoT}. 

External radiation acting solely by heating particles suspended in a flow have shown to modify the global motion  and to induce turbulent thermal convection.
The work of \cite{zamansky2014radiation, zamansky2016turbulent} considered a transparent fluid with suspended inertial particles subject to a constant radiation and at local thermal equilibrium with the fluid. 
Convection induced in such a system was found to be driven by individual plumes rising out of each particle with turbulent kinetic energy being the largest in the presence of a strong particle preferential concentration where the plumes of individual particles are reinforced by one-another due to their spatial proximity. 
This eventually led to a sustained turbulent thermal convection, albeit with the temperature of the system constantly increasing due to the permanently applied incident radiation. 

Internally heated convection (IHC) -- induced and sustained by the application of a bulk heating term in a fluid -- has also been studied as an idealised theoretical model by \cite{Lohse2021IHCscaling}. 
They consider a uniformly heated domain with the top and bottom walls kept at the same constant temperature. In this scenario, the bulk attains a stationary temperature depending on the strength of the heating and other parameters such as gravity or the height of the domain, while the fixed temperature boundaries works as a sink of heat, ensuring that the  temperature does not increase indefinitely.

The study of fluid systems where the heating in the bulk rather than boundary forcing is the dominant mode of thermal forcing has important implications for several natural systems. 
For example, in the mantle of the earth, the radiogenic heating from the decay of radioactive elements plays a significant role in addition to the heat transfer from the hotter inner core \citep{lay2008core}. 
The atmosphere of Venus which contains a high amount of sulphurous gases absorbs a large part of the incoming solar radiation, making this the dominant mode of heat transfer \citep{tritton1975internally} in contrast to the earth where the majority of the radiation is absorbed by the land surface and in-turn forces the atmosphere. 
The mantle of Venus is driven in large part by internal heating \citep{limare2015microwave}. 
Finally, in industrial applications, chiefly in the interior of liquid-metal batteries, convection due to internal heating is of crucial importance \citep{kim2013liquid}.

In this study we set up a "theoretical experiment" to study the possibility of controlling the global properties of a thermal flow by applying temperature fluctuations locally along particle trajectories. In our proposed idealisation, the particles are equipped with a hard-wired active protocol capable of releasing or absorbing heat by setting the temperature of each Lagrangian, thermally active tracer as a function of the local velocity field of the underlying fluid background. Our system is internally heated/cooled by these virtual particles so that the average heating term $\Phi$ is statistically zero and hence the average temperature attained by the fluid is unchanged by the forcing. 
The heat injection by the particles is the only energy source for the  system, since the horizontal boundaries are periodic and the top and bottom walls are adiabatic. 
The aims of the set-up are multifold. First, as a proof of concept, we wish to demonstrate that it is possible to invent hard-wired Lagrangian protocols that can cause global flow transitions. Second, we hope to trigger more studies using phenomenological or data-driven approaches to achieve control of complex systems. Finally, by acting on thermal plumes, we hope one can better understand their role in determining the organisation of the global flow. 


The remainder of the article is organised in the following manner. 
In Section~\ref{sec:Methods}, we introduce the model equations for the system, the particle temperature protocol and describe the numerical experiments conducted. 
In Section~\ref{sec:results}, we present and discuss our main findings from the numerical experiments and finally in Section~\ref{sec:Concl}, we present our conclusions as well as possible future directions for further investigation. 

\section{Methods}\label{sec:Methods}
\begin{figure}
    \centering
    \includegraphics[width = \columnwidth]{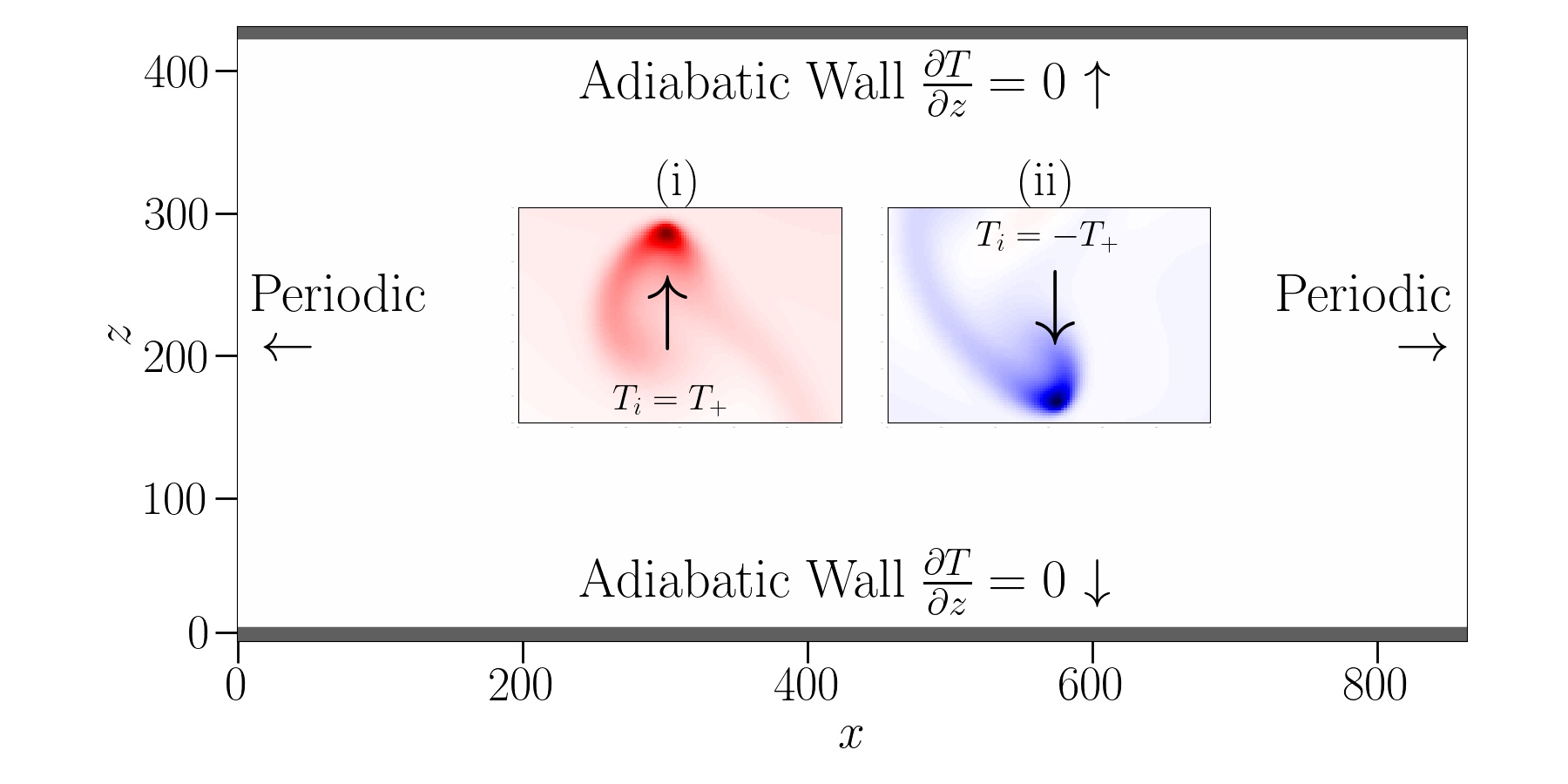}
    \caption{An overview of the methods applied in the study. 
    The domain consists of adiabatic walls at the top and bottom while the lateral boundaries are periodic. 
    In inset (i) and (ii), we show a rising hot particle with temperature $T_+$ and a falling cold particle with temperature $-T_+$ respectively. 
    }
    \label{fig:methods}
\end{figure}

The protocol for particle forcing is as follows: virtual tracer particles are initially randomly placed in a 2D region of length $L_x$ and height $L_z$ with a fluid at rest. 
The initial temperature of the fluid is set to an unstable configuration with warmer temperatures at the bottom of the domain and colder temperatures at the top of the domain. 
The particles are idealised to have an infinite heat capacity and a temperature determined by an imposed protocol in which rising particles moving vertically upward are warm with a positive temperature $T_+$, while the temperature of falling particles is set to $-T_+$ (see figure~\ref{fig:methods}) so the average temperature of the fluid remains constant. 
The temperature of the fluid near the particle relaxes to the temperature of the particle at a rate proportional to the difference between the local fluid temperature $T$ and the particle temperature $T_p$, with a relaxation time $\tau = 1/\alpha$. 

\subsection{Fluid Equations of Motion}
The fluid velocity $\uu = (u,v)$ and temperature $T$ follow the equations 
\begin{gather} 
 \nab \cdot \uu = 0, \label{eq:incomp} \\
  \partial_t \uu + (\uu \cdot\nab) \uu = - \nab p + \nu \nabla^2 \uu - \beta T \boldsymbol{g}, \label{eq:Nav-Stokes} \\
  \frac{\partial T}{\partial t} + \uu \cdot \nab T = \kappa \nabla^2 T - \sum_{i=1}^{N_p} \Bigl(\alpha_i(\br,t) \bigl[T(\br,t) - \Tp_i(t)\bigr]\Bigr).  \label{eq:Heat-eqn}
\end{gather}
where \eqref{eq:incomp} and \eqref{eq:Nav-Stokes} are the incompressible Navier-Stokes equations for a fluid with unit density and average temperature $T_0=0$ with a buoyancy-force term according to the Boussinesq approximation, where the density variations are small and enter the equations only via the gravity-force term. 
Here $p$ is the fluid pressure, $\nu$ is the kinematic viscosity, and $\beta$ is the thermal expansion coefficient. 
Temperature is advected and diffused by Equation \eqref{eq:Heat-eqn} where  $\kappa$ the thermal conductivity and the last term on the rhs is a heat source term (i.e., a thermal forcing) that depends on the  particles (see later). 

The domain is periodic in the horizontal $x$-direction while the top and bottom walls at $z=0$ and $z=L_z$ are adiabatic with $\uu = 0$, that is
\begin{gather}\label{eq:R-B_BoundaryCs}
   \partial_z T |_{z=0} =  \partial_z T |_{z=L_z} = 0,\\
   \boldsymbol{u}(z=0) = \boldsymbol{u}(z=L_z) = 0.
\end{gather}
Note that the only source of energy injected into the system is the heat supplied by the particles.

\subsection{Equations of Particle Motion}
Each particle is assumed to be a point-like, thermally active tracer. 
The $N_p$  particles with positions $  \{\rp_1,\rp_2,\dots,\rp_{N_p}\}$ and temperatures $\{ \Tp_1, \Tp_2, \dots , \Tp_{N_p}\}$ follow the local fluid velocity
\begin{equation}\label{eq:tracer}
    \frac{d \rp_i}{dt} = \uu(\rp_i(t),t).
\end{equation}

To mimic an effective particle size, concerning its thermal properties, we imagine that each particle exerts a thermal forcing on the fluid in its immediate vicinity up to a cut-off distance $\eta$. 
The feedback of the particle is defined as a local heat injection term proportional to the difference between the underlying fluid temperature, at the location of the particle, and the instantaneous particle temperature. 
Furthermore, to have a smooth thermal forcing, we assume that the strength of the coupling $\alpha_i$ (with dimension inverse of time) between the $i$-th particle and the fluid at time $t$ and position $\br$ has the form of a Gaussian with a peak at the particle location $\rp_i(t)$ (see inset (ii) of Fig. \ref{fig:methods}), given by
\begin{equation} \label{eq:Gaussian}
    \alpha_i(\br,t)= 
    \begin{cases}
        \alpha_0 \,\exp{\Bigl(-\frac{|\br - \rp_i(t)|^2}{2c^2} \Bigr)}, & \text{ if } |\br - \rp_i(t)| \le\eta,\\
        0, & \text{ if } |\br - \rp_i(t)| > \eta.
    \end{cases}
\end{equation}
Here, $\alpha_0$ is the coupling strength at the particle location and $c$ is the size of the virtual particle (referred to as particle size). 
In fact, $c$ determines the sharpness of the peak of the Gaussian function $\alpha_i$: the Gaussian peaks more sharply and falls off more quickly for smaller $c$. On the other hand, $\eta$ is simply a cut-off length for the thermal forcing by the particle. 
While setting a cut-off length introduces a discontinuity in the system, we note that for the typical parameters chosen in the study, the value of $\alpha$ at the cut-off length is $\approx 0.011 \alpha_0$. This value is only $1$\% of the value at $\br = \br_i$. 
The cut-off is used to save computational resources and to avoid calculating vanishingly small values of the Gaussian curve for points far away from a particle (large values of $|\br - \br_i|)$.

The thermal forcing due to the $i$-th particle $\Phi_i$ at location $\br$ is
\begin{equation}
    \Phi_i(\br,t) = - \alpha_i (\br,t) \bigl[T(\br,t) - \Tp_i (t) \bigr],
\end{equation}
and the total thermal forcing at a given location $\br$ due to all $N_p$ particles reads
\begin{equation}
    \Phi(\br,t) = - \sum_{i=1}^{N_p} \Bigl(\alpha_i(\br,t) \bigl[T(\br,t) - \Tp_i(t)\bigr]\Bigr).
\end{equation}

Here $\Phi(\boldsymbol{r},t)$ is of exactly the same form as eq.~\eqref{eq:Heat-eqn}. It defines the relationship between the particle positions and the thermal forcing at a given location in the fluid, and thus can be directly substituted into the heat-equation, which is an Eulerian description of the temperature.

To summarise, each particle influences a fixed region surrounding itself and when two particles are within distance $2\eta$, their thermal effects are additive in the overlapping region. The coupling between the particle and the fluid is two-way, since the particle actively heats or cools the fluid, while the fluid acceleration due to this thermal forcing also accelerates the particle. 

\subsection{Particle Temperature Policy}
The temperatures of the particles are determined by a binary policy where the $i$-th particle has either a positive value $T_+$ or a negative value $-T_+$ depending on the sign of the vertical velocity of the particle $v_i(t)$:
\begin{equation}
    \Tp_i = 
    \begin{cases}
        T_+ , &  \text{if } v_i > 0,\\
       -T_+ , &\text{if } v_i < 0.
    \end{cases}
\end{equation}
Since the particle is a tracer, $v_i$ is the same as the vertical velocity of the fluid at the particle location $v(\rp_i,t)$. 
$T_+$ is a parameter that sets the temperature scale of the system. 
By heating the upward moving fluid regions and conversely, cooling the downward moving regions, this policy should enhance thermal convection by amplifying any updrafts or downdrafts if they exist. 
Particles are coupled to each other via their effects on the fluid and because of the flow thermal diffusivity.  

Our policy leads to a sharp discontinuity in the particle temperature when the particle changes direction. 
Furthermore, the temperature would rapidly fluctuate between $T_+$ and $-T_+$ at the top and bottom walls where the velocity is very small and the flow is mainly horizontal. To ensure that this doesn't affect our results, we verified that setting  $T_i = 0$ for particles within one grid length from the top and bottom walls, where the vertical velocity of the particle fluctuates rapidly from small positive values to small negative values, leads to (statistically) the same flows. 
We have also verified that all results reported below are robust against small change of the above protocol, e.g. by setting a threshold velocity $v_0$ such that $T_i=0$ when $|v|<v_0$. 


\begin{table}
    \centering
    \begin{tabular}{ >{\centering}m{0.09\textwidth} >{\centering}m{0.04\textwidth} >{\centering}m{0.04\textwidth} >{\centering}m{0.06\textwidth}>{\centering}m{0.06\textwidth}>{\centering}m{0.09\textwidth}>{\centering}m{0.14\textwidth}>{\centering}m{0.09\textwidth}>{\centering}m{0.06\textwidth}>{\centering}m{0.04\textwidth} >{\centering\arraybackslash}m{0.06\textwidth} }
        \hline
        Parameter & $L_x$ & $L_y$& $\nu$& $\kappa$ & $\alpha_0$ & $T_+$ & $g$ & $c$ & $\beta$ & $N_p$  \\ 
        \hline
         Range & 864 & 432 & $\frac{1}{1500}$ & $\frac{1}{1500}$ & $[0.0001-0.005]$  & $[2.5 \times 10^{-8} - 0.05]$ & $8 \times 10^{-6}$ & $[0.5 - \sqrt{2}]$ & 1 & $[48 - 960]$  \\
         \hline
    \end{tabular}
    \caption{List of parameters used in the study along with the range of values in simulation units. }
    \label{tab:param_range}
\end{table}




\subsection{Numerical Experiments}
The fluid equations \eqref{eq:incomp}--\eqref{eq:Heat-eqn} are solved by the Lattice-Boltzmann method (see Appendix~\ref{sec:AppA_NumMeth} for details), together with the particle evolution as a tracer given by equation~\eqref{eq:tracer}. 
The particle evolution is solved by the two-step Adams-Bashforth method. 
We start from an initially unstable vertical temperature profile of 
\begin{equation}\label{eq:initTprof}
    T(z) = T_+ \tanh \Bigl( \frac{L_z}{2} - z \Bigr) .
\end{equation}
The two-way coupled particle-fluid system is evolved until the flow reaches a statistically stationary kinetic energy independent of the initial conditions for the flow velocity, temperature  and particles positions. 
All measurements and analyses are performed at this steady state for different sets of parameters, varying $T_+$, $N_p$, $\alpha_0$, and $c$. 
The cut-off distance for the particles $\eta$ is kept constant throughout the study. 

All results presented in this study are for a 2D fluid domain resolved with 864 grid points in the horizontal direction and 432 grid points in the vertical direction. 
With the Lattice Boltzmann grid spacing $\Delta x = 1$, we have $L_x = 864$ and $L_z = 432$. 
The particles have a fixed cut-off distance $\eta = 3$ in computational units, while their size $c$, is varied. 
$\alpha_0$ is varied from $10^{-4}$ to $5 \times 10^{-3}$ in simulation units.
The temperature $T_+$ is varied over several orders of magnitude. 
All temperatures in this study are reported in units of $T_s/0.025$ where $T_s$ is the temperature in simulation units. 
Thus, $T = 0.1$ corresponds to a temperature of $T_s = 0.0025$ in simulation units. 
This convention is chosen solely to make it easier to compare the scales of the various $T_+$ and make the manuscript more readable. 
The values of the parameters are summarised in table~\ref{tab:param_range}. 

In order to have dimensionless quantities, we define a \textit{typical velocity} $u_0$, given by
\begin{equation}\label{eq:u_0}
    u_0 = \sqrt{c g \beta \frac{\alpha_0}{\alpha_0 + \frac{\kappa}{2c^2}}T_+},
\end{equation}
where $c$ is the size of the particle as defined in equation \eqref{eq:Gaussian}. 
The  form (\ref{eq:u_0}) was suggested by studying the evolution of single particles experiments  at varying $\alpha_0$ and $c$, where the  rms value of the vertical particle velocity  was found to scale as in  \eqref{eq:u_0}. 
In particular, we find that the particle velocity statistics remain independent of the domain height $L_z$, justifying the choice of $c$ as the length scale of the system. 
The fluid near the particle relaxes to the temperature of the particle, and this relatively hotter/cooler local plume rises/falls. 
The tracer particle in turn responds to the fluid and accelerates at a rate that depends on the temperature anomaly, gravity $g$ and $\beta$. 
This is similar to other thermal flows such as Rayleigh-B\'enard convection. 
The local heating is high when $c$ is large because a wider region around each particle is thermally forced. 

In the given protocol, the temperature scale alone is set by the particles and this is in turn sets a natural velocity field for the system. Given that the particles are active and the fluid-particle coupling is highly non-linear,  it would be a hard task to estimate other quantities associated with thermal systems usually defined in the literature such as a Richardson number or a Brunt-Valsala frequency as this would require a-priori knowledge of the average state of the system.

The quantity 
\begin{equation}\label{eq:T_a}
    T_a = \frac{\alpha_0}{\alpha_0 + \frac{\kappa}{2c^2}} T_+
\end{equation}
is interpreted as an \textit{effective temperature} reached in the vicinity of each particle.
The empirical prefactor $\alpha_0/(\alpha_0 + \frac{\kappa}{2c^2})$ by which $T_+$ is multiplied is a constant that gives the rate of relaxation of the fluid temperature to the particle temperature compared with the rate at which heat is diffused away from the particle by conduction, which is proportional to $\kappa/c^2$.  When $\alpha_0\to0$, then $T_a\to0$, because the fluid is no longer coupled to the particle and there is no energy input to the system. 
When $\alpha_0 \gg \kappa/c^2 $, then $T_a\to T_+$, meaning the fluid  attains the local particle temperature. 
For large $\kappa$, the heat is rapidly conducted away from the particle so that  effective temperature  is lower, where again $T_a\to 0$ when $\kappa\to\infty$ while the case of small $\kappa$ is similar to that of large $\alpha_0$. 
In our study, $\alpha_0$ and $\kappa/c^2$ are of comparable magnitude. 

Furthermore, we define the normalized \textit{turbulent kinetic energy} $E_k(t)$ of the system as
\begin{equation}\label{eq:normTKE}
    E_k(t) = \frac{1}{2} \frac{\bigl\langle |\uu(t)|^2 \bigr\rangle_V}{u_0^2 N_p},
\end{equation}
where $\langle \cdot \rangle_V$ represents the average over the entire domain at a given time. 
We also define with an overline $\overline{E}_k$ as the average normalized turbulent kinetic energy (TKE), i.e. 
\begin{equation}\label{eq:avgnormTKE}
    \overline{E}_k = \bigl\langle E_k(t) \bigr\rangle_t,
\end{equation}
where $\langle\cdot\rangle_t$ denotes the time average after the flow reaches a statistically stationary regime. 
If the particles are sparse and their motion is independent of each other, the kinetic energy of the system would simply be a sum of the motion of the individual particles and we would expect $\overline{E}_k$ of the flow to attain a constant value. However, if the motions of the particle are not independent of each other, the variation of $\overline{E}_k$ as a function of $N_p$ is not possible to predict a-priori.

\section{Results}\label{sec:results}

\subsection{Stable and Convective Configurations}
\begin{figure}
    \centering
    \includegraphics[width = \linewidth, keepaspectratio]{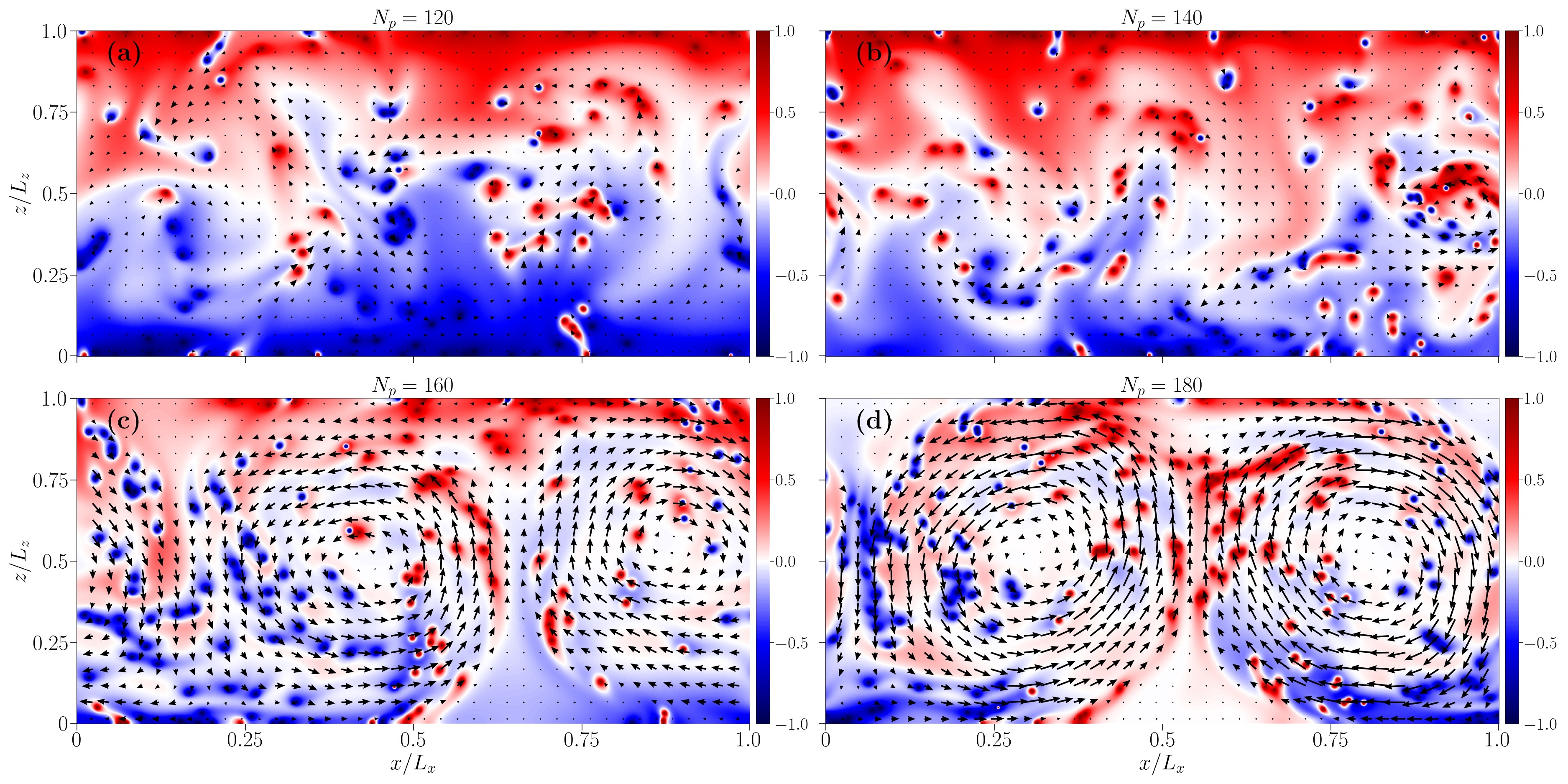}
    \caption{Snapshots of the temperature field $T(\br,t)/T_+$ at a given instant of time for $T_+ = 0.1$, $\alpha_0 = 0.005$, $c=1$ and at changing  $N_p = 120,140, 160,180$ in panels (a), (b), (c) and (d), respectively. 
    The colour palette varies from red to blue where red indicates $T = T_+$ and blue indicates $T = -T_+$. 
    The black arrows show the velocity field with the length of the arrow representing the relative magnitude of the velocity with identical scaling for all four panels. 
    The top panels show a stable  configuration while the bottom panels show a convective  configuration. } 
    \label{fig:Tsnap_comp}
\end{figure}

First, we vary the number of virtual particles $N_p$. Figure~\ref{fig:Tsnap_comp} shows
four cases, where we visualize snapshots of the temperature and velocity fields. Thereby the rising particle temperature $T_+$, particle-fluid coupling strength $\alpha_0$ and particle size $c$ are fixed. 
The figure indicates that there are two distinct stationary typical  configurations. 
The first, which we term \textit{stable}, is shown in the top panels (a) and (b) of  figure~\ref{fig:Tsnap_comp}. In this state, kinetic energy is low and  large scale circulation is absent. 
Particles are either nearly still and close to the top and bottom walls or they execute a slow vertical motion independently one from the others, propelled by their higher or lower temperature compared to the bulk. 
When the particle concentration reaches beyond a certain threshold,
the individual thermal effect of the particles aggregates and triggers a transition to a second state shown in the bottom panels (c) and (d) of figure~\ref{fig:Tsnap_comp}. 
This convective state enjoys a large scale circulation, the  presence of  rising and  falling plumes with the  particles trajectories synchronized with the  large-scale recirculation regions. 
In figure~\ref{fig:Tsnap_comp}, this transition occurs for $N_p \sim 150$.

\begin{figure}
    \centering
    \includegraphics[width = \linewidth, keepaspectratio]{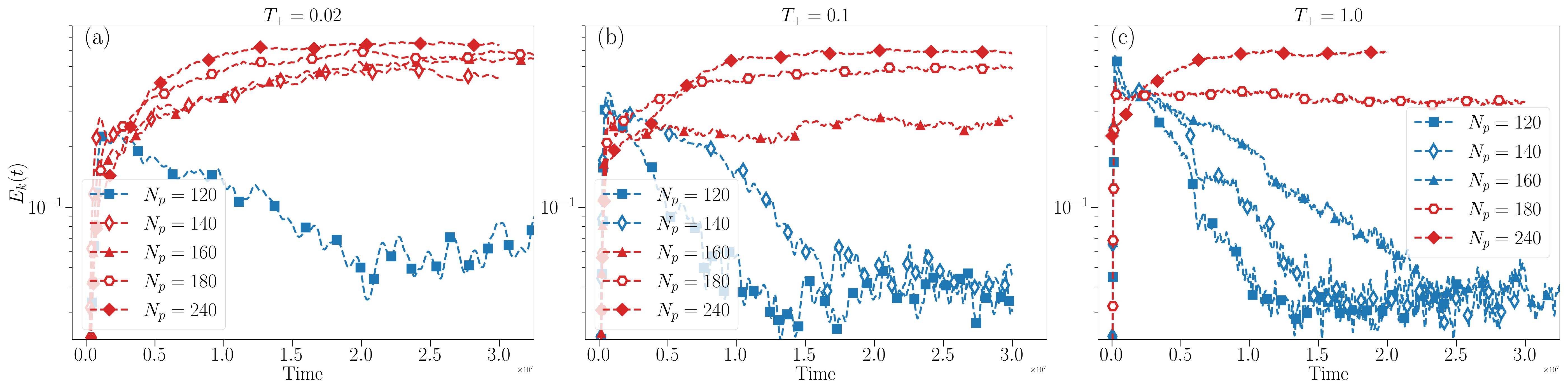}
    \caption{Time evolution of $E_k(t)$ for flows with (a) $T_+ =0.02$, (b) $T_+0.1$ and (c) $T_+=1.0$ with $\alpha_0 = 0.005$ and $c = 1$ kept fixed. 
    Stable configurations are plotted in blue while convective  configurations are plotted in red. 
    The time is in simulation time units.}
    \label{fig:ke_evol_comp1}
\end{figure}

In figure~\ref{fig:ke_evol_comp1} we show the time evolution of the TKE for parameters before and after the transitions.  Panels (a), (b) and (c) corresponds to  $T_+ = 0.02$, $T_+ = 0.1$ and $T_+ = 1.0$, respectively, with $\alpha_0$ and $c$ remaining fixed. 
The blue curves represent stable configurations while the red curves represent convective configurations. 
The kinetic energy first increases due to the unstable temperature gradient imposed on the initial condition. 
At later times, the thermal forcing by  the tracers is dominant and the flow attains a statistically stationary kinetic energy where $E_k(t)$ either shows a large value (red curves), corresponding to a convective flow shown  qualitatively in figure~\ref{fig:Tsnap_comp} or a low value (blue curves) corresponding to a quasi  stable flow. 

Two further points are note-worthy about the transition from figure~\ref{fig:ke_evol_comp1}. 
Firstly, the transition is abrupt:it is enough to add  very few particles to hvae a jump $\gtrsim 5$ in the normalised kinetic energy. 
It should be noted that the expression of $E_k(t)$ is normalized by $N_p$ in the denominator, so the absolute increase in kinetic energy is even greater. 
Secondly, the critical  $N_p$ depends slightly on $T_+$, where for larger $T_+$, the transition occurs at a slightly larger $N_p$. 
We see that in panel (a) with $T_+ = 0.02$, the transition lies between $N_p = 120$ and $N_p = 140$ while in panel (c) with $T_+ = 1.0$, the transition lies between $N_p = 160$ and $N_p = 180$, with the case of $T_+ = 0.1$ in panel (b) showing an intermediate behavior.
This weak dependence on $T_+$ which will be further commented upon later. 
It has been verified that the transitions are robust by replacing the initial unstable profile with an initial temperature field of $T=0$ everywhere with particles being either hot or cold with probability $0.5$ each. 

\begin{figure}
    \centering
    \includegraphics[width = \linewidth, keepaspectratio]{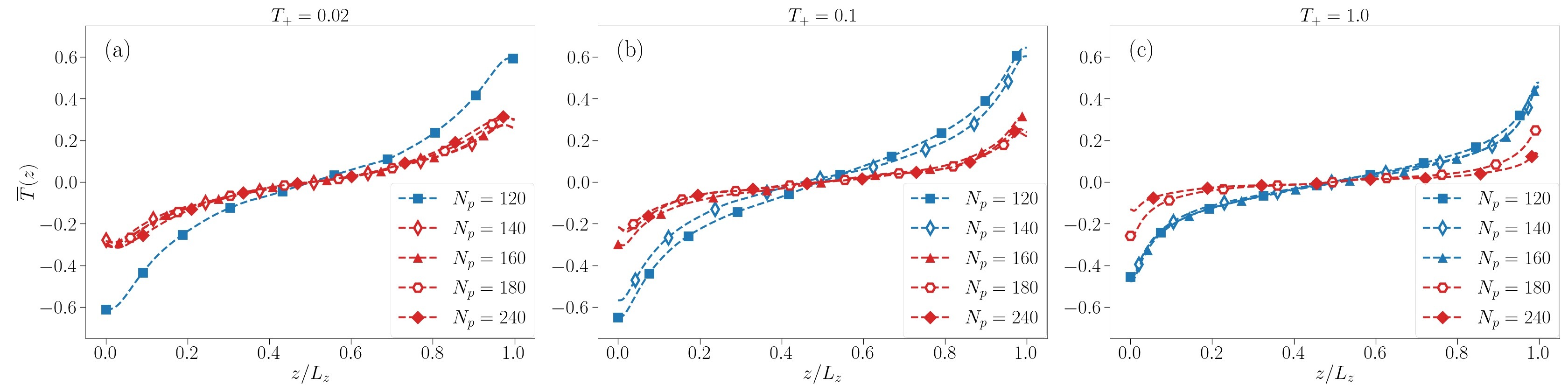}
    \caption{Time-averaged vertical temperature profile divided by $T_+$ plotted against the vertical height for various $N_p$ close to the transition $N_p$ for $T_+ = 0.02$ (a), $T_+ = 0.1$ (b) and $T_+ = 1.0$ (c). 
    Stable configurations are plotted in blue while convective configurations are shown in red.}
    \label{fig:Tprof_comp}
\end{figure}

In figure~\ref{fig:Tprof_comp}, we show a comparison of the normalised time-averaged vertical temperature profiles $\overline{T}(z)$ for the same set of flows given by
\begin{equation}
    \overline{T}(z) = \frac{\bigl\langle T(\br,t) \bigr\rangle_{x,t}}{T_+},
\end{equation}
where $\langle \cdot \rangle_{x,t}$ represents the time-average at a given height $z$. 
Notice that the temperature gradients for the stable flows (blue) show a strongly stable profile ($\partial_z T > 0$) while the convective flows still show an overall stable temperature profile but with weaker gradients so that the temperature difference between the top and the bottom adiabatic walls are much smaller. 
In the presence of a large-scale circulation, the temperature field is more effectively transported and mixed throughout the domain. 
We also see that with increase in $T_+$, the convective configurations show a flatter temperature profile for the corresponding $N_p$ of lower $T_+$ flows, i.e., for example, the red curves in panel (c) are much flatter than those in panel (a).

The dual-nature of the effect of the virtual particles is observed here -- the particles tend to make the flow more stable by carrying heat away from the lower half of the domain while carrying heat towards the upper half of the domain.
Thus, the larger $T_+$ is, the more stable the system becomes.
However, when a certain threshold of particles is reached, the situation changes -- the virtual particles together create a persistent large-scale flow and now the convection is strong enough to overcome the stable temperature gradient. 

\begin{figure}
    \centering
    \includegraphics[width = 0.9\linewidth, keepaspectratio]{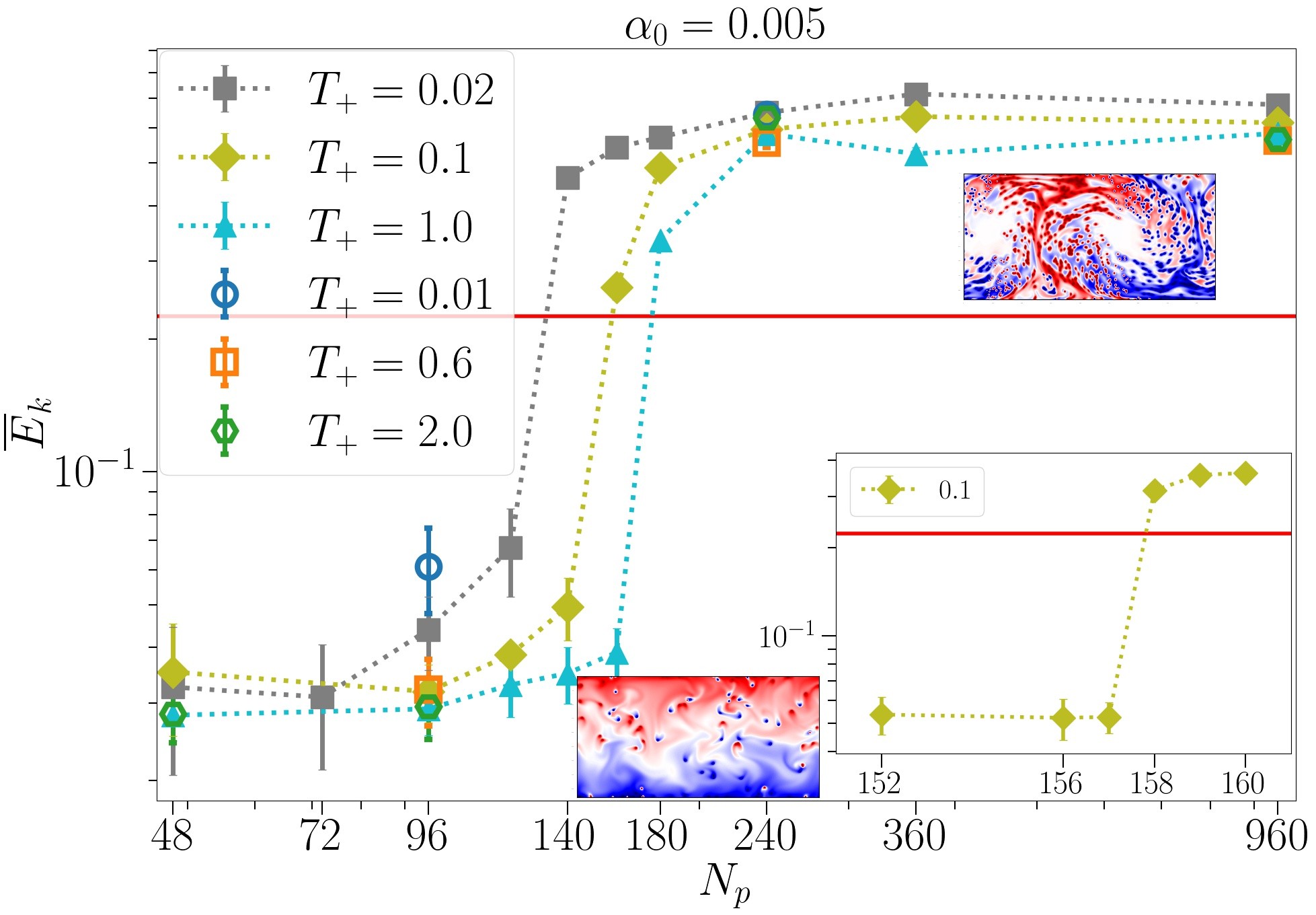}
    \caption{Time averaged normalized TKE $\overline{E}_k$ as a function of $N_p$ for various $T_+$ (shown in legend) for $\alpha_0 = 0.005$ and $c = 1$. 
    The inset (in the lower right corner) shows the behavior of a flow with $T_+ = 0.1$ very close to the transition $N_p$. 
    Also shown are instantaneous snapshots of the temperature field for a stable configuration (bottom) and a convective configuration (top right). 
    The error bars indicate the standard deviation of the temporal fluctuations of $E_k(t)$ around the average kinetic energy in the stationary regime.
    }
    \label{fig:KEVsNp}
\end{figure}

In figure~\ref{fig:KEVsNp}, we take a closer look at the transition by plotting the average normalised TKE of the flows as defined in equation \eqref{eq:avgnormTKE} against $N_p$, for the same $\alpha_0$ as above, for various $T_+$. 
Here we see that $\overline{E_k}$ initially remains constant for small values of $N_p$ and then increases rapidly during the transition from a stable flow to a convective flow. Once the convective flow has been set up, the value of $\overline{E_k}$ saturates to a higher constant value for large $N_p$.
The sharp increase of TKE at a transition $N_p$ is once again clearly visible. 
We empirically define a value of $E^0_k = 0.225$ indicated by the horizontal red line as the transition point where for stable end states, $\overline{E_k} < E^0_k$ and vice-versa for the convective end state. 
The sharpness of the transition is examined more closely in the inset of the figure for a given $T_+$. 
It is seen that the transition occurs for an increase of just one single  particle. 
The dependence of the transition on $T_+$ is weak, for $T_+$ varying over 2 orders of magnitude  the transition occurs at nearly the same $N_p$. 

\subsection{Large-scale Circulation and Heat Transfer}\label{subsec:Lsc-Heat}
While the existence of the large-scale circulation is apparent from the visualisations of the temperature and velocity fields, it is possible to infer its presence quantitatively from the fluid energy spectrum. In particular, we consider the spectrum in the horizontal direction taken at the mid-plane $z_0 = L_z/2$, given by 
\begin{equation}\label{eq:u-spec}
    E_{\uu}(k_x) = \frac{1}{2} \Bigl\langle \big|\hat{\uu}(k_x,z_0,t)\bigr|^2 \Bigr\rangle_t,
\end{equation}
 and $\hat{\uu} (k_x,z_0,t)$ are the Fourier coefficients of the field $\uu$ and $\langle \cdot \rangle_t$ denotes the time averaging.  
We denote by $E_1$ the energy contained in the first Fourier mode with wavenumber $k_x = 2\pi/L_x$, $E_2$ is used for energy of the second mode ($k_x = 4 \pi/L_x$), and so on.
Moreover, we define $E_{\rm tot}$ as the sum of the energy contained in all the Fourier modes, 
\begin{equation}
    E_{\rm tot} = \sum_{i=1}^{N_k} E_i,
\end{equation}
where $N_k$ is the Fourier mode corresponding to the smallest resolved length-scale.
The strength of the large-scale circulation with a rising plume and a falling plume can be measured by the value $E_1/E_{\rm tot}$ \citep{RBflowmodes}, which measures the fraction of energy contained in the first mode, that is the smallest wave number. 
This corresponds to a cosine mode for the velocity field in the bulk, which is a close approximation when there exist two counter-rotating vortices.
When such a large-scale flow is present, we would have $E_1/E_{\rm tot} \gg 0$, while if the flow lacks large-scale convection, we would have a flatter energy spectrum with $E_{\rm tot} \gg E_1$ and $E_1\sim E_2$. 

\begin{figure}
    \centering
    \begin{subfigure}{0.49\columnwidth}
        \includegraphics[width=\columnwidth]{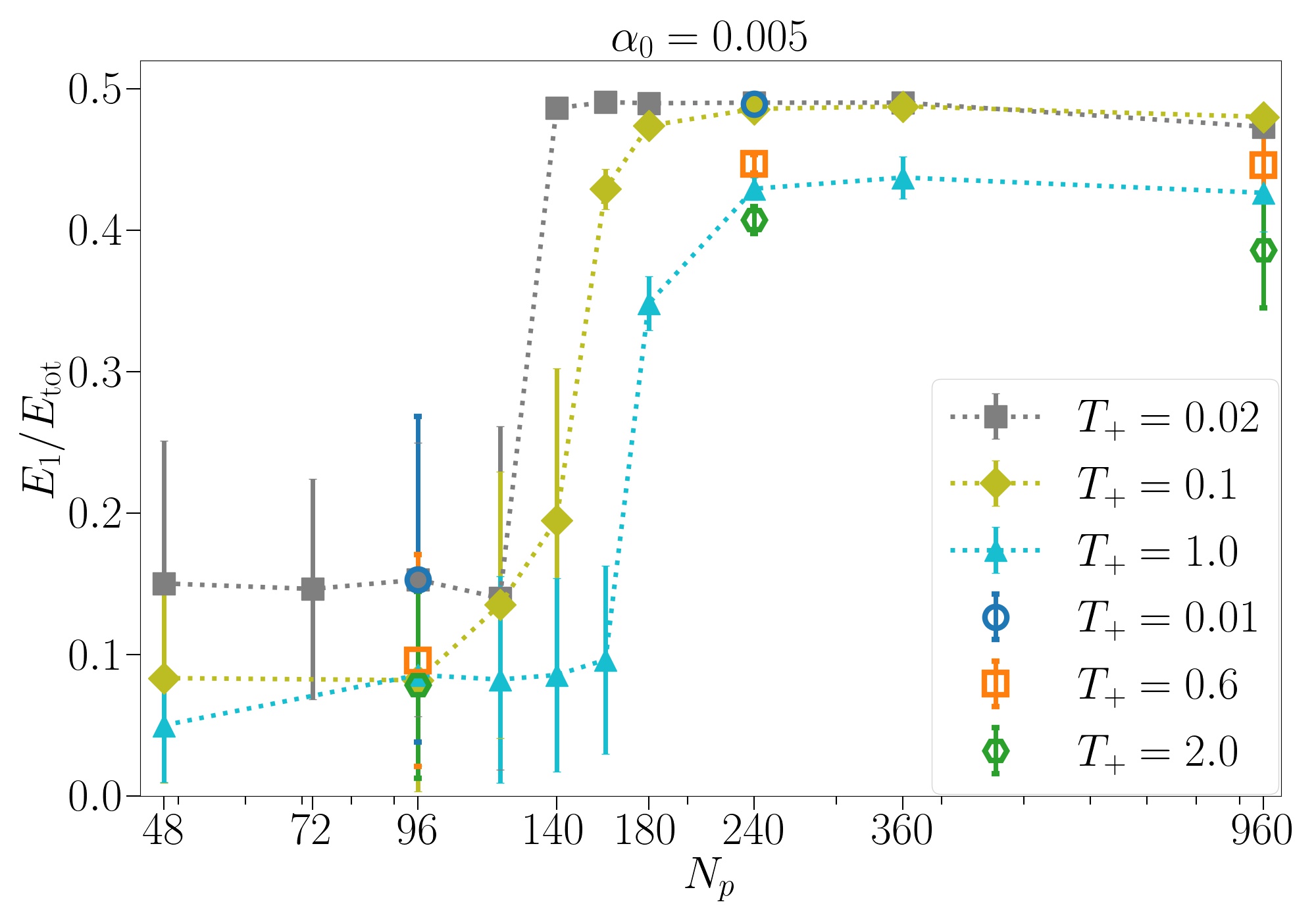}
        \caption{}
    \end{subfigure}
    \begin{subfigure}{0.49\columnwidth}
        \includegraphics[width=\columnwidth]{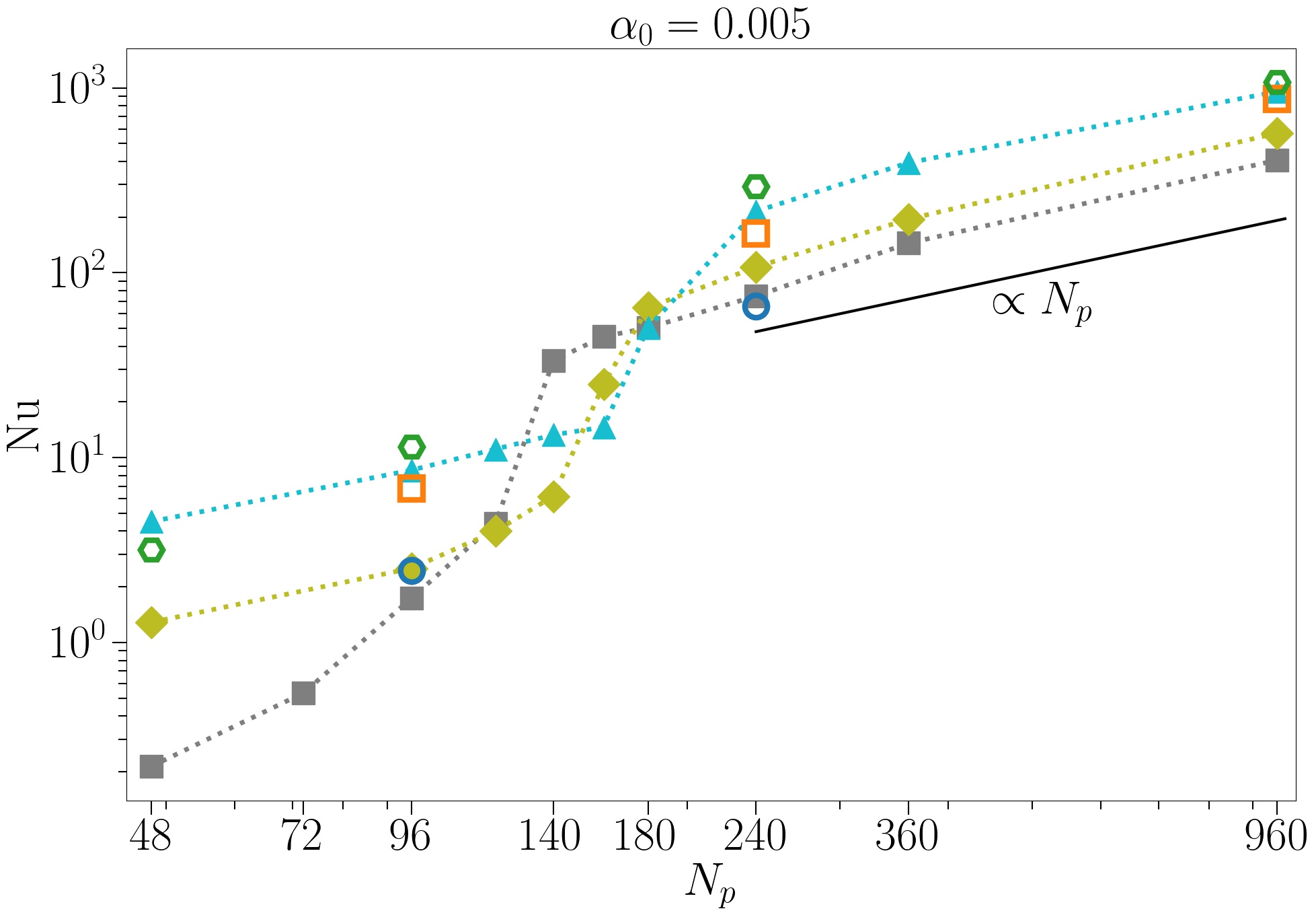}
        \caption{}
    \end{subfigure}
    \begin{subfigure}{0.49\columnwidth}
        \includegraphics[width=\columnwidth]{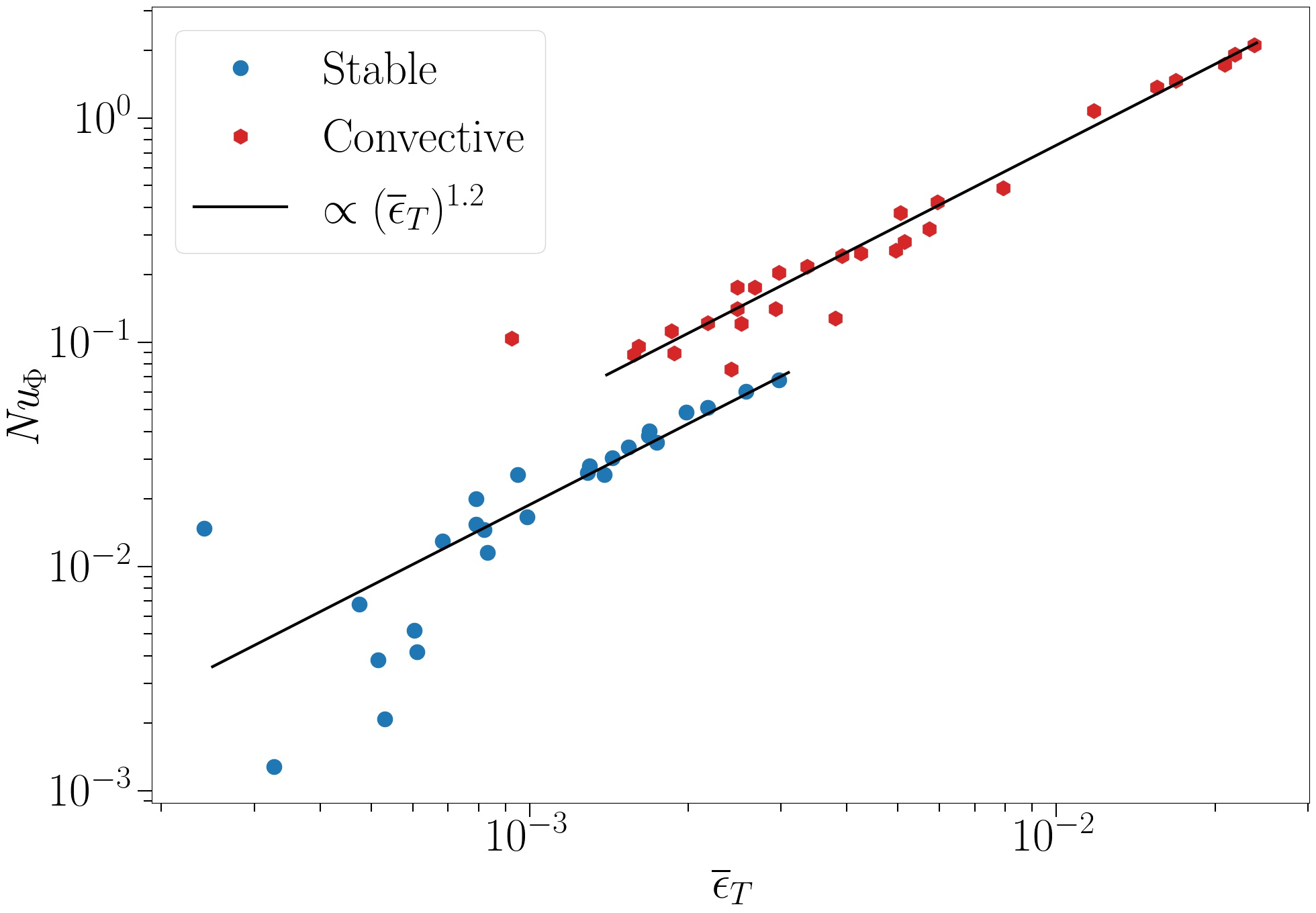}
        \caption{}
    \end{subfigure}
    \caption{(a) $E_1/E_{\rm tot}$ for varying $N_p$ for various values of $T_+$. 
    Error bars show the temporal fluctuations of $E_1/E_{\rm tot}$ (b) $\operatorname{Nu}$ for varying $N_p$ for various values of $T_+$. 
    The black solid line shows a linear scaling with $N_p$. 
    (c) Plot of the average normalised Nusselt number $\operatorname{Nu}_\Phi$ against the average normalised thermal energy injection $\overline{\epsilon}_T$ for flows with varying parameters. Stable flows are marked with blue filled circles, convective with red filled hexagons and the two black lines scale as $(\overline{\epsilon}_T)^{1.2}$. Inset of panel (c) shows the scaling of $\operatorname{Nu}_\Phi$ as a function of $N_p$ for 3 different $T_+$.}
    \label{fig:Lsc-str}
\end{figure}

In figure~\ref{fig:Lsc-str}(a), we plot the strength of the large-scale circulation $E_1/E_{\rm tot}$ for varying $N_p$. 
We see clearly here that corresponding to a jump in the magnitude of the TKE seen in figure~\ref{fig:KEVsNp}, there is also a similar large increase in the ratio of kinetic energy contained in the largest-scale. 
Given that $\overline{E}_k$ takes into account the typical velocity of a single particle as well as the number of particles, the excess kinetic energy clearly comes from the large-scale circulation that arises after the transition, a cumulative particle effect. 

Figure~\ref{fig:Lsc-str}(b) shows the dimensionless Nusselt number, $\operatorname{Nu}$, defined as 
\begin{equation}\label{eq:Nusselt}
    \operatorname{Nu} = \frac{\Bigl\langle v T - \kappa \frac{\partial T}{\partial z} \Bigr\rangle_{V,t}}{\frac{\kappa \overline{\Delta T}}{L_z}}, 
\end{equation}
where $\langle \cdot \rangle_{V,t}$ represents average over the entire domain and time, $v$ is the vertical fluid velocity and $\overline{\Delta T}$ is the time-averaged temperature difference between the top and bottom walls given by 
\begin{equation}
    \overline{\Delta T} = \left \langle T(x,L_z) \right \rangle_{x,t} - \left \langle T(x,0) \right \rangle_{x,t}.
\end{equation}
Here, the Nusselt number is defined in analogy with  Rayleigh-B\'enard convection: it is the ratio of heat transfer due to convection and the heat transfer by conduction with the difference that the temperature jump is taken in the opposite sense because of the presence of a stable mean profile. 
Due to the adiabatic boundary conditions imposed at the top and bottom walls ($\partial_z T = 0$) along with the no-slip boundary condition for the velocity ($\uu = 0$), the value of the Nusselt number is $0$ at the top and bottom walls.
Thus, the boundary walls do not contribute to the heat transfer. 
The Nusselt number naturally increases proportionally with the number of particles. 

We see in figure \ref{fig:Lsc-str}(b) that the value of $\operatorname{Nu}$ increases gradually with increase in $N_p$, followed by a large increase around the transition $N_p$ and then settling to a roughly linear increase with $N_p$ in the convective regime.
The reason for the large increase of $\operatorname{Nu}$ at the transition is two-fold. 
Firstly, the increase in TKE overall leads to an increase in the convective heat transfer which further increases $vT$. 
Secondly, with more effective mixing of the temperature and a weakly stable temperature gradient, $\overline{\Delta T}$ in the denominator also has a smaller magnitude. 

Another way to quantify the heat transfer is to divide it by the typical forcing $\Phi$ multiplied by the length-scale of the system.
This is similar to the normalisation procedure of \citep{Lohse2021IHCscaling} applied to internally heated convection (with volume forcing). 
The effective temperature $T_a$ defined in equation \eqref{eq:T_a} was introduced as a typical value of the temperature attained by the fluid in the vicinity of the particle with an associated length-scale $c$ for each particle. 
In a similar vein, $\alpha_0 (T_+ - T_a)$ can be considered the typical thermal forcing acting on the fluid. 
We use two  dimensionless response parameters of the system. 
First, we define the normalised Nusselt number $\operatorname{Nu}_{\Phi}$ given by
\begin{equation}
    \operatorname{Nu}_{\Phi} = \frac{\Bigl\langle v T - \kappa \frac{\partial T}{\partial z} \Bigr\rangle_{V,t}}{c \alpha_0 (T_+ - T_a)}. 
\end{equation}
$\operatorname{Nu}_{\Phi}$ measures the heat transfer by convection relative to the input typical thermal forcing multiplied by the length scale of the system. 

In the stationary regime, the thermal dissipation rate $\epsilon_T$ is given by $\langle \Phi T \rangle_{V,t}$ (see Appendix~\ref{sec:AppC_ThermD}) and is normalised as 
\begin{equation}
    \overline{\epsilon}_T = \frac{\langle \Phi T \rangle_{V,t}}{\alpha_0 (T_+ - T_a) T_+}.
\end{equation}
The normalisation factor is once again the typical forcing multiplied by the temperature scale. 

In figure~\ref{fig:Lsc-str}(c) we plot the normalised Nusselt number $\operatorname{Nu}_{\Phi}$ against the normalised thermal dissipation $\overline{\epsilon}_T$, quantifying the measured convective response of the fluid to the measured input thermal forcing for varying $T_+$, $c$, $\alpha_0$ and $N_p$. 
It is seen that there exists a global scaling of these two quantities for both the flow regimes, stable and convective with a rough scaling of $\operatorname{Nu}_{\Phi} \propto (\overline{\epsilon}_T)^{1.2}$. 
However, the higher magnitude of the normalised Nusselt number in the convective case differentiates it from the stable flows. In the inset of panel (c) we can see that Nu$_{\Phi}$ also shows a transition and a sharp increase in magnitude from stable to convective. 

The above findings are consistent with a situation that can be briefly described as such -- individual particles thermally coupled to the fluid have a small zone of influence and release or absorb heat in their immediate vicinity. 
Thus, each particle contributes to the thermal injection into the domain as well as the vertical heat transfer across the domain both of which increase with the increase in number of particles. 
In the stable regime, the main effect of the particles is to maintain the strongly stable temperature gradient. However, since there exists a maximum and minimum temperature that the fluid can attain ($T_+$ and $-T_+$ respectively), there is consequently a limit on the strength of the positive temperature gradient that can be set-up in the system, independent of the number of particles. Hence, increasing the number of particles eventually leads to a saturation of the stable temperature gradient while it continues to further destabilise the system. Thus, eventually at a critical $N_p$, the instability dominates the flow and leads to a transition to a fully convective flow.

At the transition to the convective regime, the development of the large-scale convective flow patterns and more turbulent flow leads to a large increase in the heat transfer relative even to the thermal energy injection, while also seeing a weaker stable vertical temperature gradient across the domain. 

\begin{figure}
    \centering
    \includegraphics[width = \linewidth, keepaspectratio]{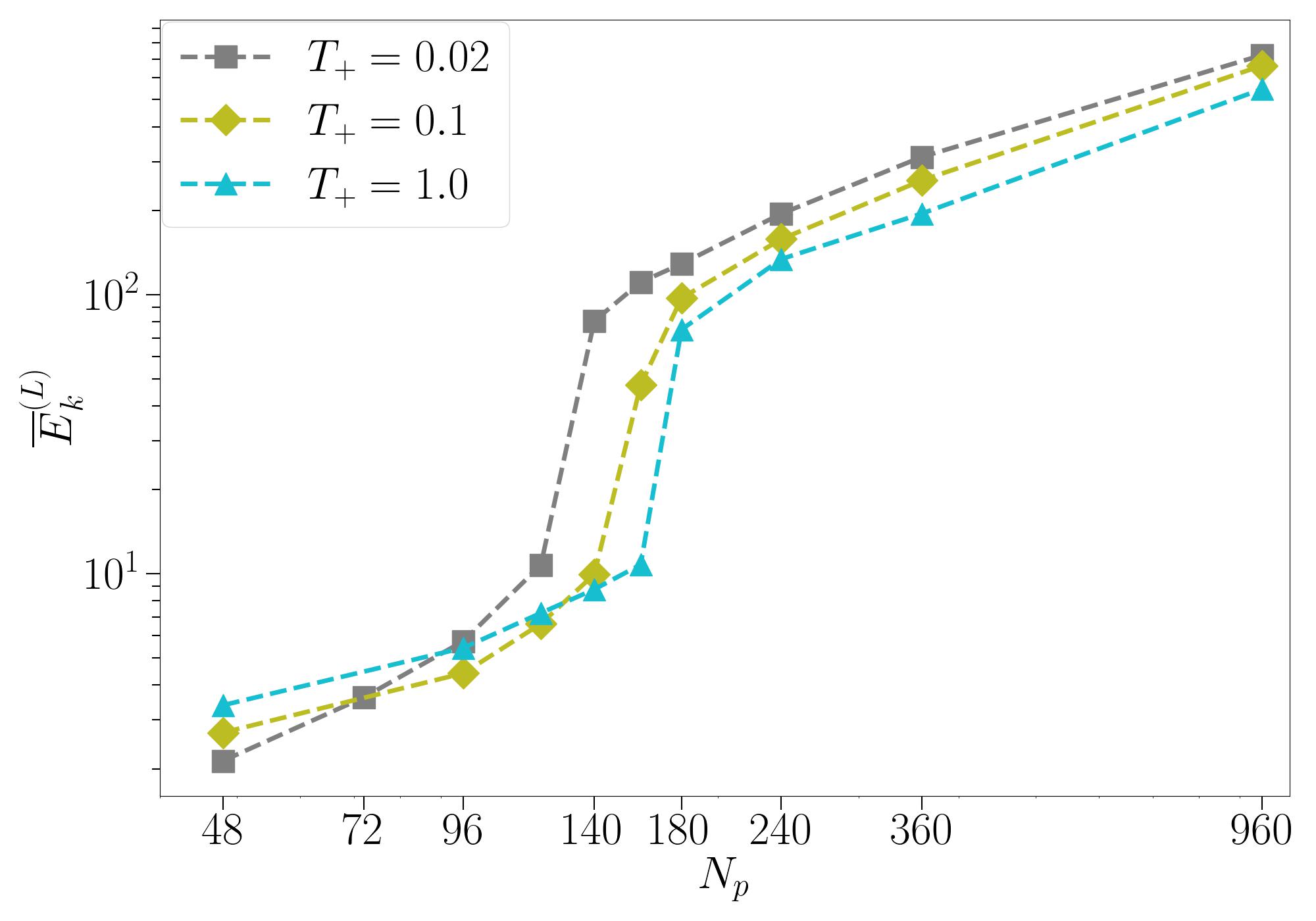}
    \caption{Average particle kinetic energy $\overline{E}_{k}^{(L)}$ plotted against $N_p$ for 3 values of $T_+$. All measurements are made with $c=1$ and $\alpha_0 = 0.005$.}
    \label{fig:prtcl_props}
\end{figure}
\subsection{Particle Kinetic Energy}
Given that the particles are not passive and are actively interacting with the fluid, their behaviour is of particular interest. We define the Lagrangian kinetic $\overline{E}_{k}^{(L)}$ as 
\begin{equation}
     \overline{E}_{k}^{(L)} = \frac{1}{u_0^2 N_p} \left \langle \sum_{i=1}^{N_p} \frac{1}{2} \left( u_i^2 + v_i^2\right) \right\rangle_t,
\end{equation}
where $u_i$ and $v_i$ are the horizontal and vertical velocity of the $i$-th particle respectively. $\overline{E}_{k}^{(L)}$ measures the average kinetic energy of a particle in the system. In figure~\ref{fig:prtcl_props}, we plot this value against $N_p$ for various $T_+$, which shows the characteristic jump at the transition to a convective flow. The absolute value of $\overline{E}_{k}^{(L)}$ is far larger than the corresponding fluid kinetic energy $\overline{E}_k$ due to the fact that the particle locations are the active regions of the flow where the energy is input into the system. Thus, in spite of being tracers, these particles selectively sample higher kinetic energy regions of the flow. $\overline{E}_{k}^{(L)}$ also shows an increase as a function of $N_p$. The reason for this behaviour is two-fold -- first, with increasing $N_p$, the fluid becomes more energetic and turbulent as discussed before. Secondly, the particles themselves are better mixed and are found more often in the more energetic regions of the fluid away from the stationary walls.  

\subsection{Comparison with Eulerian imposed thermal forcing}
\begin{figure}
    \centering
    \includegraphics[width = \linewidth, keepaspectratio]{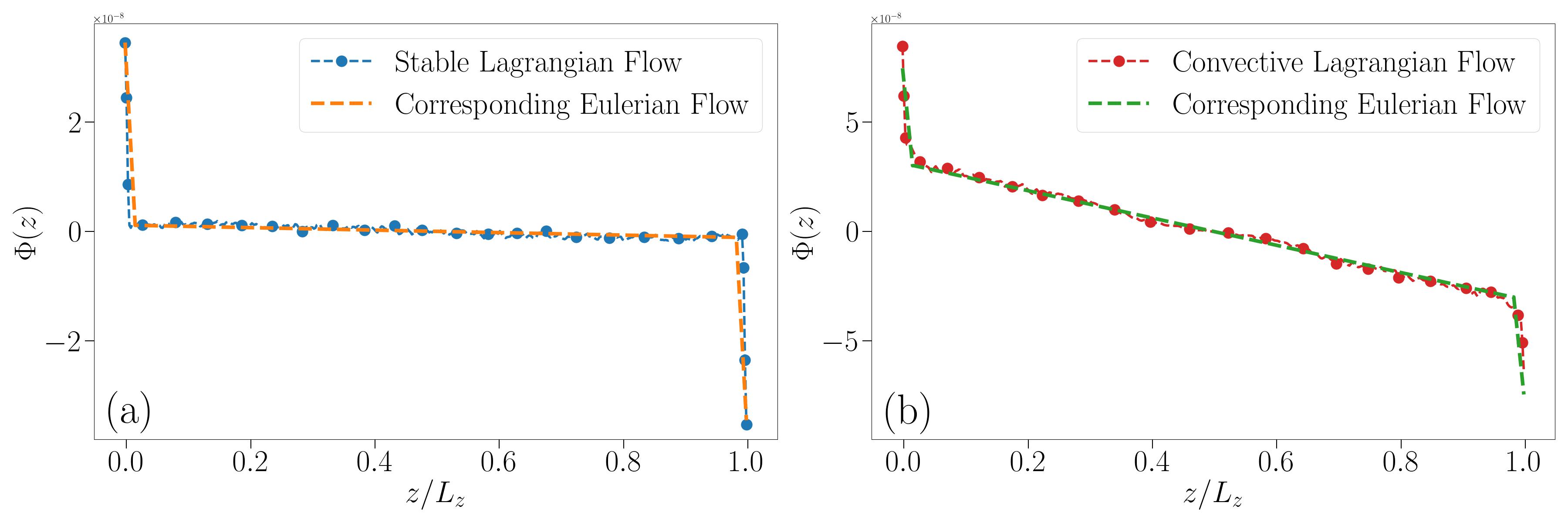}
    \caption{The measured value of the average vertical profile of thermal forcing $\Phi(z)$ for (a) a stable Lagrangian flow and for (b) a convective Lagrangian flow shown along with the profile of the corresponding applied Eulerian thermal forcing.}
    \label{fig:qprofs}
\end{figure}

We consider a thermal fluid system with a thermal forcing $\phi(z)$ uniformly applied at all times. 
The forcing is a close approximation of the vertical profile of the measured forcing $\Phi$ in the Lagrangian system as shown in figure~\ref{fig:qprofs}. 
Defining $Q(z)$, the numerator of the Nusselt number, as the average net heat transfer in the positive $z$ direction at height $z$ given by 
\begin{equation}
    Q(z) = \Bigl\langle v(\br,t) T(\br,t) - \kappa \partial_z T |_{(\br,t)} \Bigr\rangle_{x,t},
\end{equation}
where $\langle \cdot \rangle_{x,t}$ indicates the time and spatial averages  at a given height $z$, notice that averaging equation \eqref{eq:Heat-eqn} over time gives
\begin{equation}
     \Phi(z) \coloneqq \bigl\langle\Phi(\br,t) \bigr\rangle_{x,t} = \partial_z Q (z).
\end{equation}

\begin{figure}
    \centering
    \begin{subfigure}{0.49\columnwidth}
        \includegraphics[width=\columnwidth]{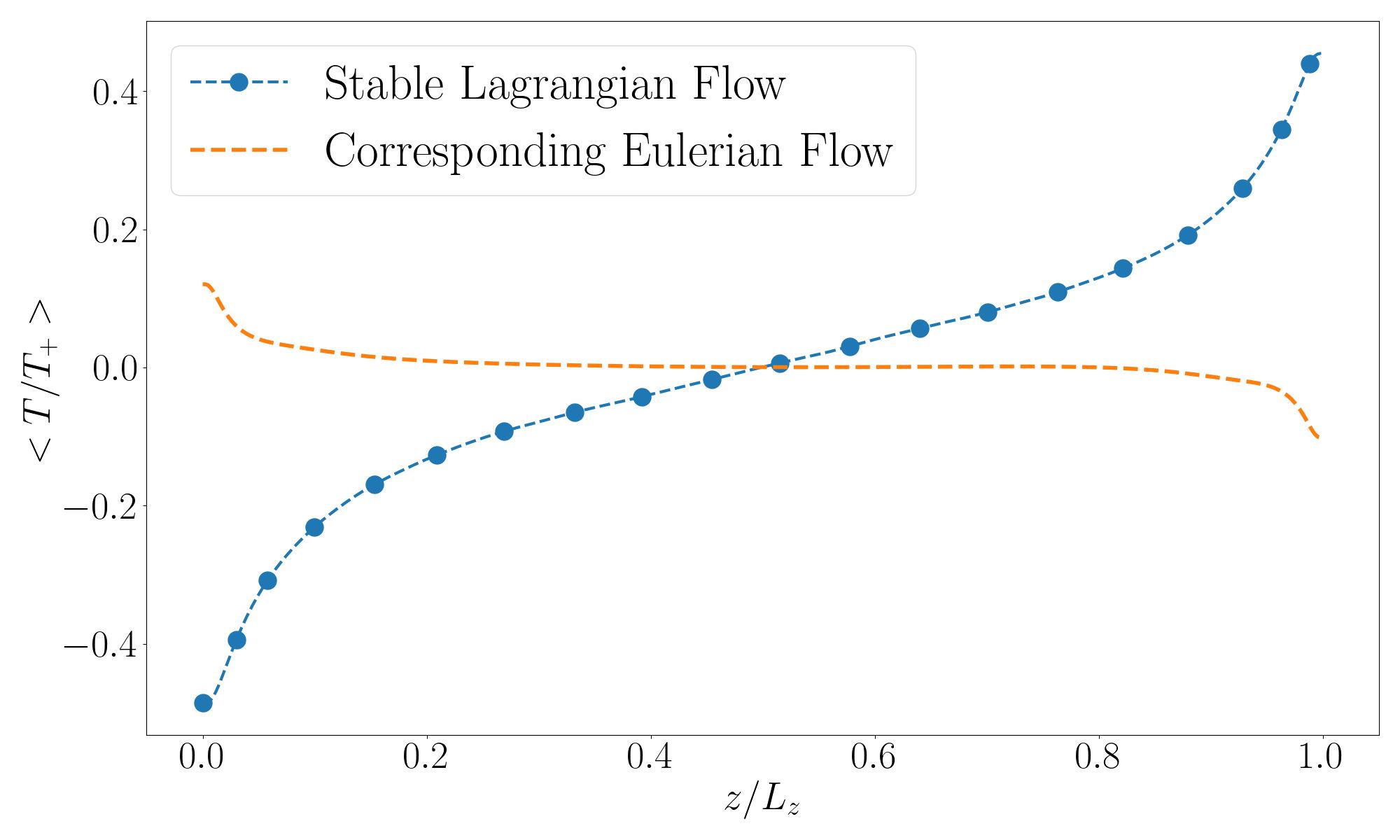}
        \caption{}
    \end{subfigure}
    \begin{subfigure}{0.49\columnwidth}
        \includegraphics[width=\columnwidth]{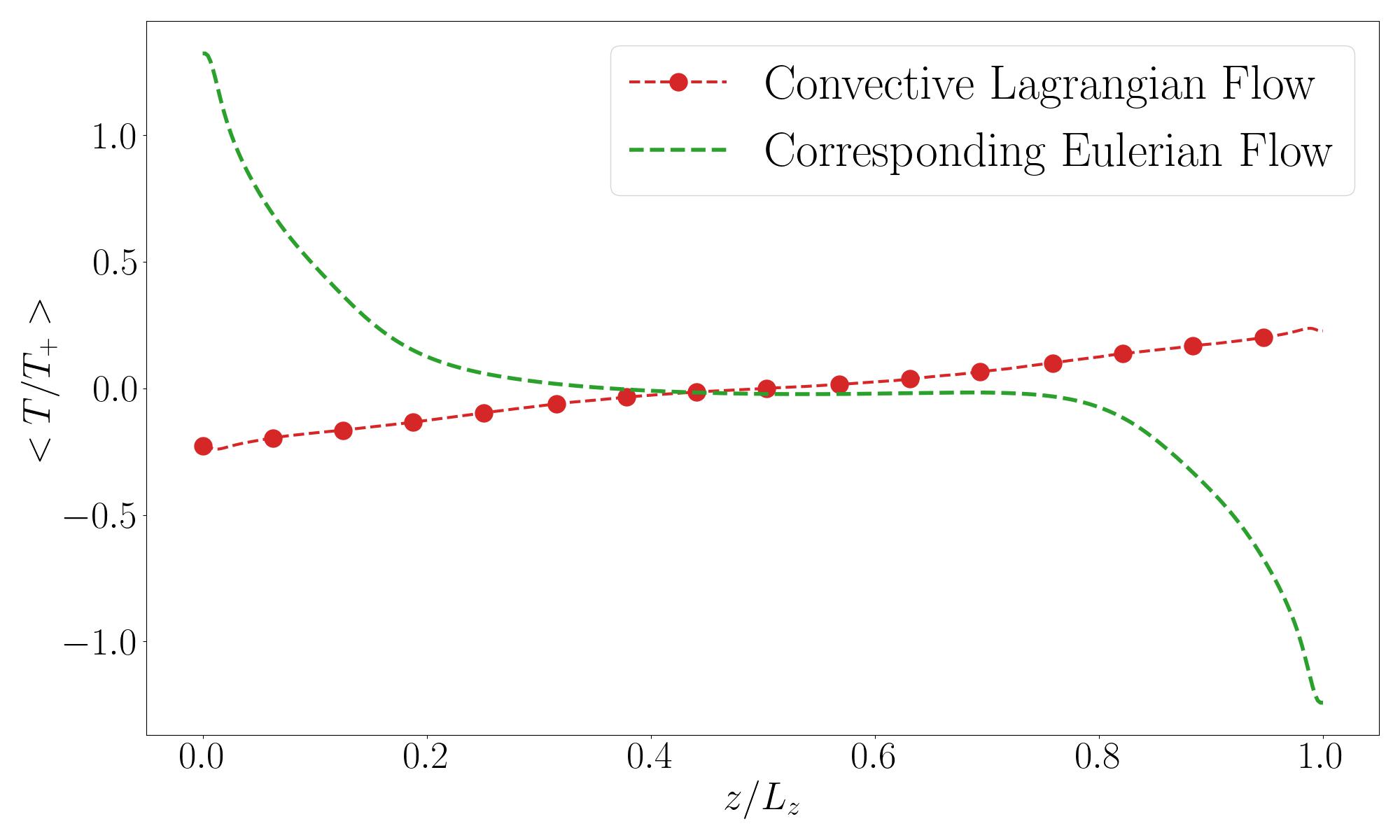}
        \caption{}
    \end{subfigure}
    \caption{(a) The normalised temperature profile for a stable Lagrangian flow (blue) compared with the measured temperature profile of a flow with an identical imposed profile of thermal forcing. 
    (b) The normalised temperature profile for a convective Lagrangian flow (red) compared with the measured temperature profile of a flow with an identical imposed profile of thermal forcing}
    \label{fig:Tprofs_Eul_Lagr}
\end{figure}

\begin{figure}
    \centering
    \includegraphics[width = \textwidth]{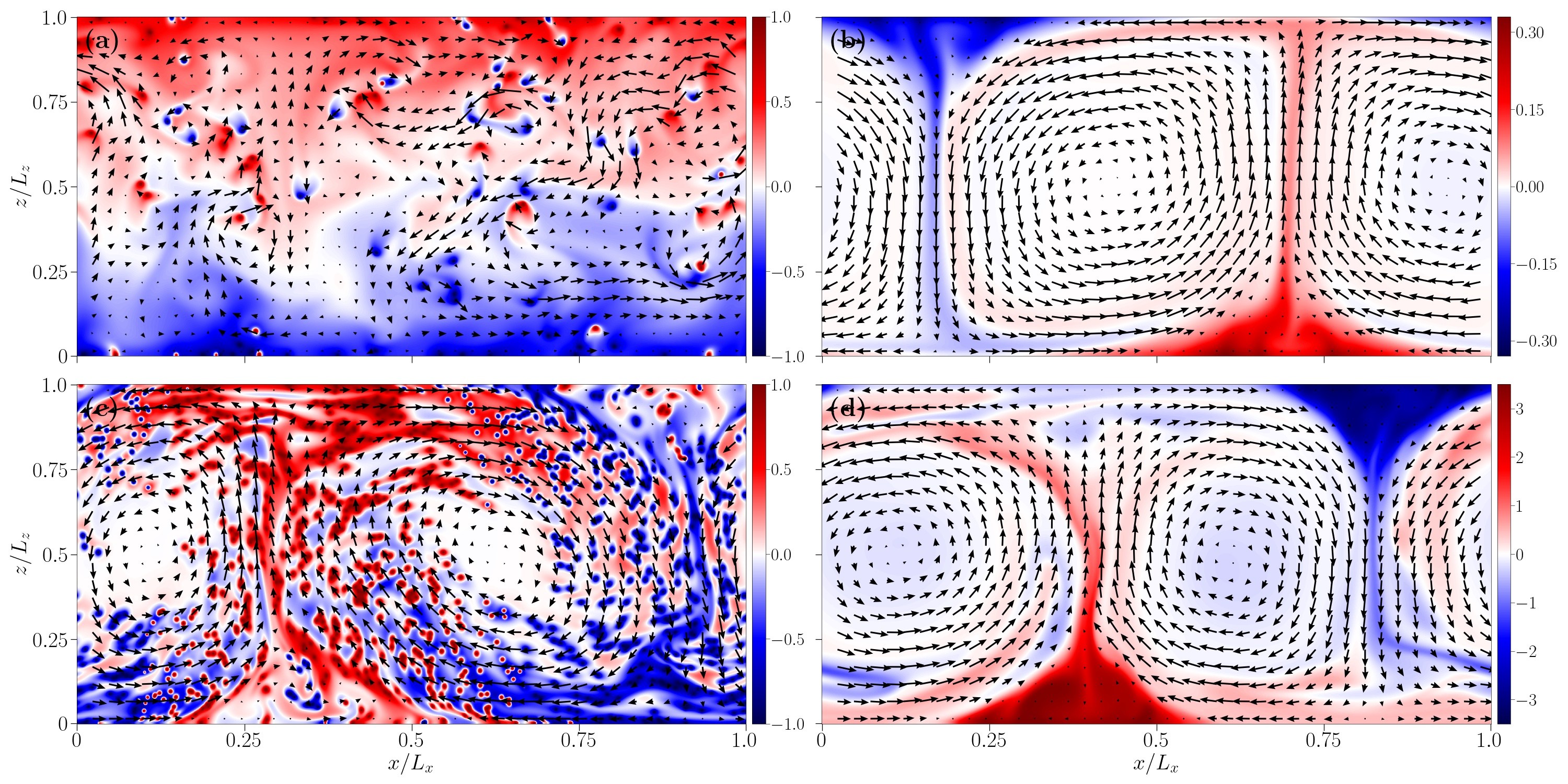}
    \caption{Snapshots of the temperature fields: (a) a stable Lagrangian flow (upper left), (c) a convective Lagrangian flow (lower left), and the two uniformly forced flows to mimic the stable (b) and convective flows (d) in right column. 
    The temperature fields $T$ are divided by the respective $T_+$. 
    The black arrows show the velocity field. 
    The length of the arrows indicate the magnitude of the velocity field within each panel -- the arrow lengths are scaled differently for different flows to allow for the clearest viewing of the flow structure. }
    \label{fig:Eul_Lagr_snaps}
\end{figure}

The comparison is made for  one stable and one convective flow. Given identical vertical profiles of thermal forcing (see figure~\ref{fig:qprofs}), one would expect that the resulting temperature profile and hence the nature of the flows would remain identical. 
However, as shown in figure~\ref{fig:Tprofs_Eul_Lagr}, the temperature profiles show a dramatic difference, with the Eulerian flows showing an unstable temperature profile similar to the Rayleigh-B\'enard Convection. 
Further, as shown in figure~\ref{fig:Eul_Lagr_snaps}, even when the thermal forcing matches the measured value from a stable configuration, the Eulerian flow with uniform thermal forcing shows a convective behavior with clear, well-defined hot and cold plumes and an unstable temperature gradient. 
Even in the convective Lagrangian case, the corresponding Eulerian flow is convective. 
Thus, the presence of the stable temperature gradients and the two  typical  configurations outlined previously is not a result of the net thermal forcing applied on the system but of the particular Lagrangian nature of the thermal tracers and the two-way coupling with the fluid. 

While it is possible to obtain a stable flow by imposing a comparatively very small value of thermal forcing in the Eulerian case, it should be emphasised here that the Lagrangian forcing is a measured quantity and not a quantity that can similarly be set independently. The thermal forcing in the Lagrangian case is not known a-priori due to the non-linear, 2-way coupling between the particles and the fluid. 

\subsection{Anomalous Behavior for Small \texorpdfstring{$T_+$}{Lg}}
\begin{figure}
    \centering
    \begin{subfigure}{0.49\columnwidth}
        \includegraphics[width=\columnwidth]{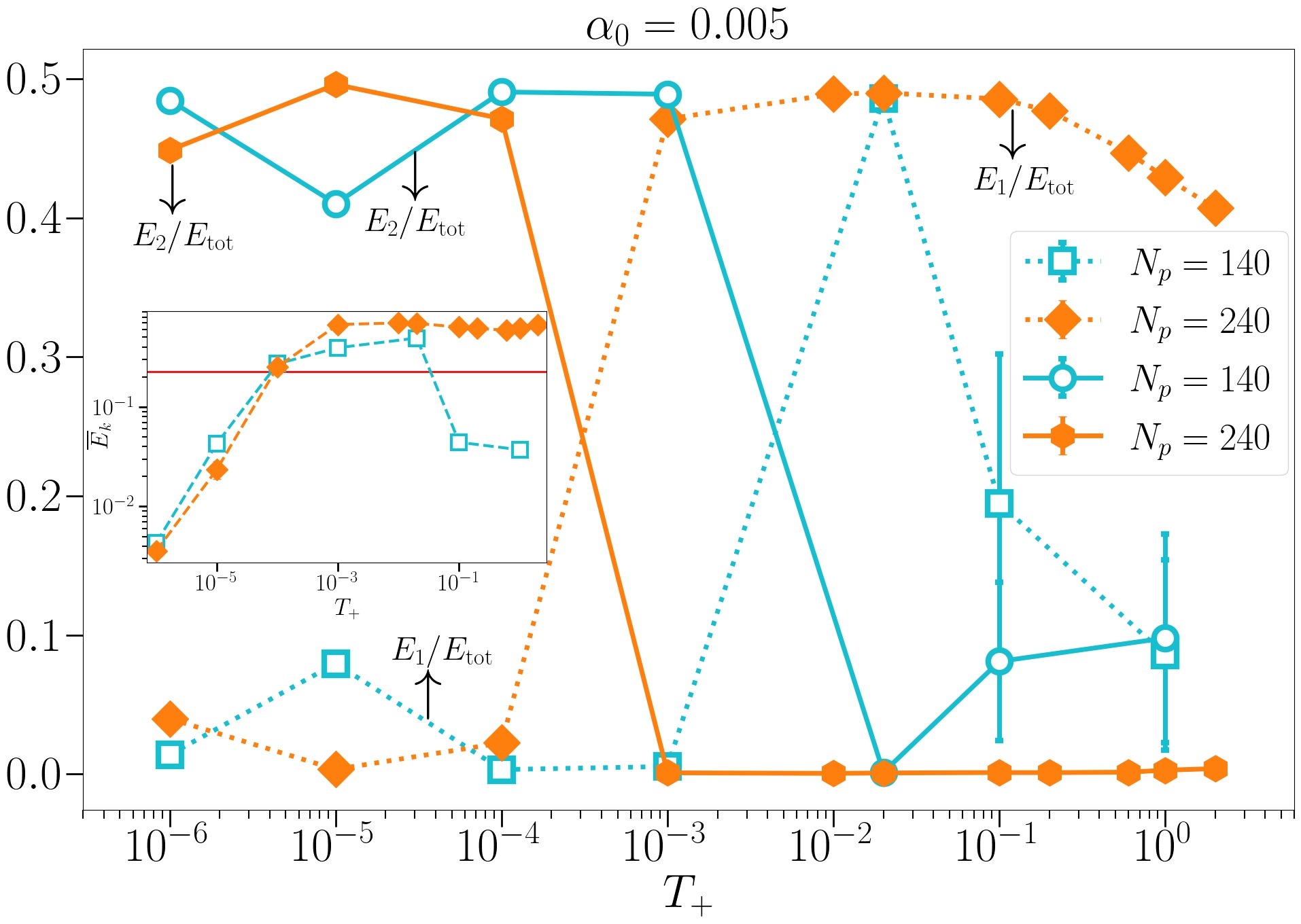}
        \caption{}
    \end{subfigure}
    \begin{subfigure}{0.49\columnwidth}
        \includegraphics[width=\columnwidth]{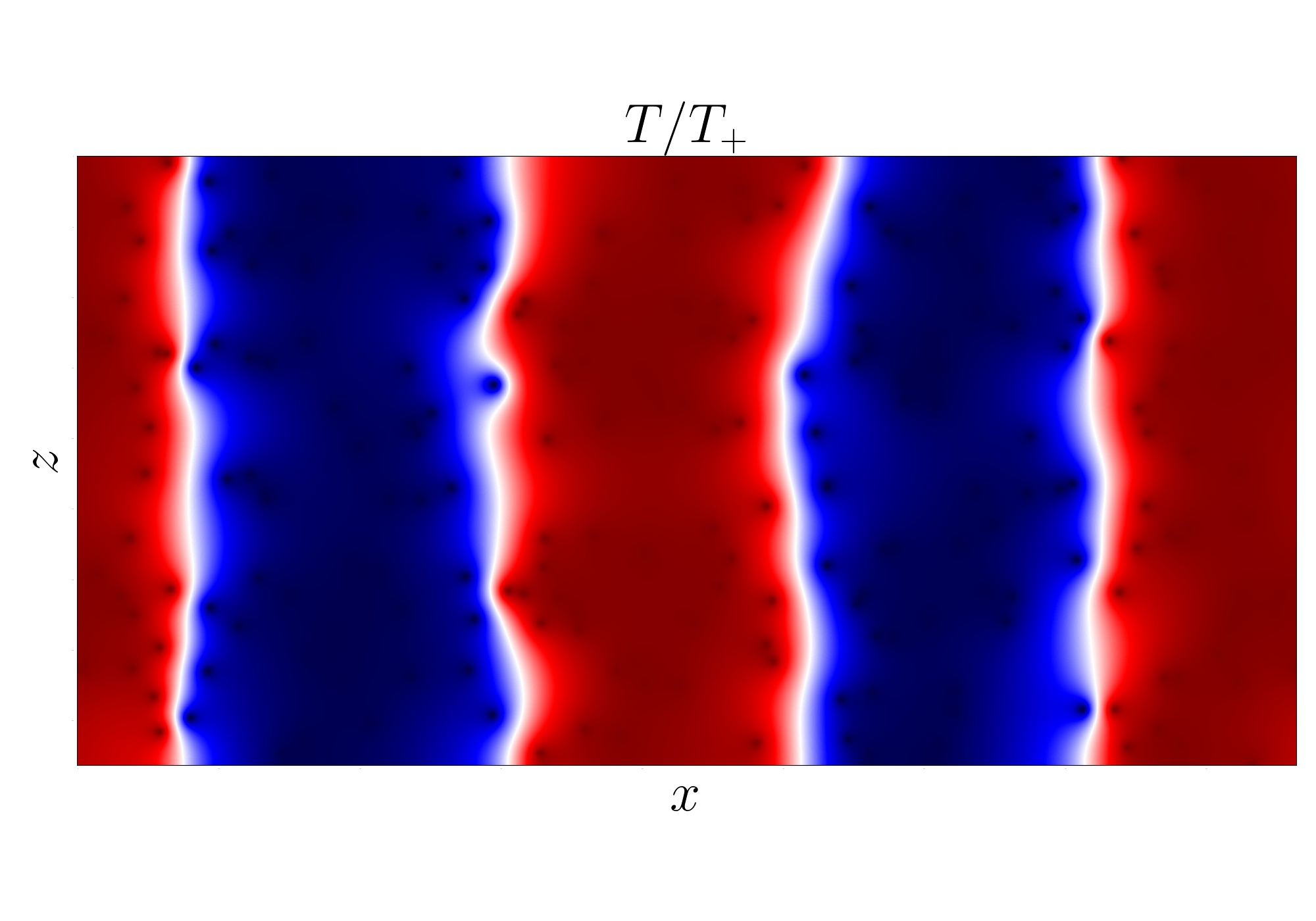}
        \caption{}
    \end{subfigure}
    \caption{(a) The ratio of kinetic energy contained in the first Fourier mode $E_1/E_{\rm tot}$ (dashed lines) and $E_2/E_{\rm tot}$ (solid lines) to the total energy contained in all modes for $N_p = 140$ and $N_p = 240$ plotted against $T_+$. 
    Inset shows the averaged normalised TKE $\overline{E}_k$ for the same parameters and the horizontal line represents $\overline{E}_k = E_k^0$. 
    (b) A snapshot of the temperature field for a columnar flow with $N_p = 240$ and $T_+ = 10^{-5}$. The colour palette varies from red to blue where red indicates $T = T_+$ and blue indicates $T = -T_+$.  }
    \label{fig:small_Tp}
\end{figure}

We have already noted in previous sections that there is weak dependence of the transition of the system on the value of $T_+$. 
In particular, it was observed that for larger $T_+$, the transition occurs at a larger $N_p$ and the stable configurations for larger $T_+$ have relatively flatter temperature gradients. 
One would conclude then that for any given $N_p$, there exists a $T_+$ small enough such that the system is convective. 
However, at very small $T_+$, the system attains a third columnar state where the temperature profile is still stable ($\partial_z T>0$) and the system has a weak convective flow (see snapshot in figure~\ref{fig:small_Tp}~(b)).

In figure~\ref{fig:small_Tp}~(a), we plot the fraction of energy contained in the first Fourier mode ($E_1/E_{\rm tot}$) as well as the second Fourier mode ($E_2/E_{\rm tot}$) to understand the large-scale behavior of the flow. 
We can see clearly that for smaller $T_+$, the second mode dominates the kinetic energy while the energy contained in the first mode approaches $0$.
This is the case until a transition $T_+$, where now the flow turns convective from columnar, with a dominance of $E_1$.
At larger $T_+$ for $N_p = 240$ (orange, filled symbols), we see that while $E_2/E_{\rm tot}$ remains small, the value of $E_1/E_{\rm tot}$ shows a decreasing trend. 
This is because as $T_+$ is increased, the flow becomes more turbulent and  small-scale velocity  features begin to appear, increasing the energy contained at higher modes. 
For $N_p = 140$ (cyan, empty symbols), the flow is columnar for $T_+ \lesssim 10^{-3}$ and transitions to convective at $T_+ \sim 0.02$, as evidenced by the values of $E_1/E_{\rm tot}$ and $E_2/E_{\rm tot}$. 
However, on increasing $T_+$ further, the flow again moves to a stable configuration, as evidenced by the fact that $E_1 \sim E_2$ which indicates the lack of any large-scale velocity flow. 
This transition is due to the effect already observed, that for increasing $T_+$, the $N_p$ of transition from stable to convective is greater.

The inset of figure~\ref{fig:small_Tp}(a) shows the normalised TKE plotted for the two given $N_p$ and varying $T_+$. 
Notice that at small $T_+$, when the flow is columnar, it is characterised by a smaller normalised TKE and kinetic energy smoothly approaches $0$ as $T_+ \to 0$.

\section{Conclusions and Discussion}\label{sec:Concl}
We have performed numerical simulations of an idealized non-isothermal 2D fluid system under the Boussinesq approximation with suspended  tracer particles. 
The particles act as heat sources or sinks depending on their vertical velocity. 
The particles are coupled to the fluid only thermally, the fluid is forced only by the action of the particles. 
Individually, each particle aids in the transport of heat away from the bottom of the domain towards the top of the domain, thus working to create a thermally more stable system. 
However, under certain conditions, the cumulative effect of the particles overpowers the tendency towards stability and the result is a system with a large-scale convective flow pattern with increased turbulent kinetic energy, larger heat transfer across the domain, maximum energy in the largest Fourier modes and a (weakly) stable vertical temperature gradient.
The main parameters of the system are the temperature of the hot, rising particles $T_+$, the number of particles $N_p$, the strength of the thermal coupling between the fluid phase and the particles $\alpha_0$ and the size of the particle $c$. 
Increasing $N_p$, $c$ and $\alpha_0$ makes the flow increasingly convective while increasing $T_+$ weakly contributes to making the flow more stable. 

This Lagrangian protocol is compared with a system with a uniform thermal forcing identical to the measured Lagrangian forcing and it is found that the temperature profiles of the Eulerian system is unstable rather than stable and a convective flow always develops. 

Extension to 3D set-ups and to cases with larger domain and/or a larger number of particles to study whether the intensity of turbulence can be increased indefinitely would also be interesting.\\

Independently of the possibility to realize a protocol like the one we studied here in a realistic experimental set-up, out study is meant to gain a new  insight about the impact of Lagrangian control on turbulent convection.
A real-world example would be a cloud of droplets moving along with an updraft -- the droplet remains uniformly hotter than the surroundings due to condensation of water onto its surface and similarly, falling cloud droplets constantly lose water to the atmosphere thus remaining cooler while moving downward. 

Our study also opens several further interesting avenues for investigation including -but not limited to- the formulation of similar  protocols where the properties of the suspended particles is optimized by a data-driven approach  to attain complex controls and modulation of fluid flows. 
It would also be very interesting to explore other protocols where the particle properties are coupled with the underlying dynamics of the fluid local to the particle, for example where the temperature of the particle is proportional to its vertical velocity as well as cases where the momentum exchange of the particle with the fluid is also considered. Whether other protocols could also exhibit a transition like the one observed in this study and the physical reasons for the presence or absence of a transition may also be a question for future researchers.

Other avenues for further research include a dynamic comparison between the convective system described in this study driven by Lagrangian forcing with the Rayleigh-B\'enard convection as well as a statistical comparison between the trajectories of the active tracers with the trajectories of tracers suspended in a RB convective system. These could provide new insights into the behaviour of thermal convection.

\section*{Funding}
This project has received funding from the European Union's Horizon 2020 research and innovation programme under the Marie Sklodowska-Curie grant agreement No 765048. 
This work was supported by the European Research Council (ERC) under the European Union’s Horizon 2020 research and innovation programme (Grant Agreement No. 882340).\\
\\
\textbf{Declaration of interests}
The authors report no conflict of interest

\section*{Data Availability Statement}
Data available on request from the authors -- The data that support the findings of this study are available from the corresponding author upon reasonable request.

\appendix

\section{Numerical Methods}\label{sec:AppA_NumMeth}

\subsection{Lattice Boltzmann Method}
The fluid equations are solved by the Lattice Boltzmann method with two sets of populations using a standard D2Q9 grid.
\begin{align}\label{LB_equations}
        f_i(\br+\boldsymbol{c}_i \Delta t, t + \Delta t) &= f_i(\br,t) - \frac{f_i - f^{\rm eq}}{\tau_f} \Delta t + S_i \Delta t, \\
        g_i(\br+\boldsymbol{c}_i \Delta t, t + \Delta t) &= g_i(\br,t) - \frac{g_i - g^{\rm eq}}{\tau_g} \Delta t + q_i \Delta t.
\end{align}
The evolution of the two sets of populations $f$ and $g$, representing the fluid and the thermal phase respectively, 
follow the Lattice Boltzmann equations with a Bhatnagar–Gross–Krook (BGK) collision operator. 
The vectors $\boldsymbol{c}_i $ for $i=1,\dots,9$ are the discrete particle velocities, $\Delta t$ is the lattice time-step, 
so that $\boldsymbol{c}_i \Delta t$ go from each lattice point to the 8 nearest neighbouring lattice points in the uniform 2D grid and $\boldsymbol{c}_0 = 0$. 
$S_i$ and $q_i$ represent the momentum forcing (buoyancy) and the thermal forcing respectively. 
The time-step and the grid spacing respectively $\Delta t = \Delta r = 1$, as is the standard practice. 
$f^{\rm eq}$ and $g^{\rm eq}$ are the equilibrium population distributions as defined in \cite{HEetal_2population} given by 
\begin{align}
    f^{\rm eq} &= w_i \rho \Bigl( 1+\frac{\uu\cdot \boldsymbol{c}_i}{c^2_s} + \frac{(\uu\cdot \boldsymbol{c}_i)^2}{2c^4_s} - \frac{\uu\cdot \uu}{2c^2_s}\Bigr), \label{eq:feq}\\
    g^{\rm eq} &= w_i T \Bigl( 1 + \frac{\uu\cdot \boldsymbol{c}_i}{c^2_s} + \frac{(\uu\cdot \boldsymbol{c}_i)^2}{2c^4_s} - \frac{\uu\cdot \uu}{2c^2_s}\Bigr),
\end{align}
where $w_i$ are the weights for each population set by the grid used, D2Q9 in this study. 
$c_s$ is the lattice speed of sound set by the choice of $\boldsymbol{c}_i$. 

$\tau_f$ and $\tau_g$ are respectively the fluid and the thermal relaxation times which set the values for kinematic viscosity $\nu$ and thermal conductivity $\kappa$ as 
\begin{align}
        \nu &= c_s^2 (\tau_f - 0.5), \\
        \kappa &= c_s^2 (\tau_g - 0.5).
\end{align}
To account for the buoyancy force term, the Guo-forcing scheme \citep{Paper:Guo-forcing} is employed with 
\begin{equation}
    S_i = \Bigl( 1 - \frac{\Delta t}{2 \tau_f} \Bigr) w_i \Bigl( \frac{\boldsymbol{c_i} 
    - \uu}{c^2_s} + \frac{(\boldsymbol{c_i} \cdot \uu) \boldsymbol{c_i}}{c^4_s}  \Bigr) \cdot \boldsymbol{F},
\end{equation}
where $\boldsymbol{F}$ is the physical force vector. 

The fluid hydrodynamic quantities at each point in space and time are obtained from the various moments of the populations as
\begin{align}
    \rho &= \sum_i f_i,  \\
    \uu &= \frac{1}{\rho} \sum_i f_i \boldsymbol{c}_i + \frac{\boldsymbol{F}}{2\rho}. \label{eq:LB_uhdydro}
\end{align}
The ease of implementation of the Guo-forcing scheme is from the fact that the velocity $\uu$ that enters the expression for $f^{\rm eq}$ in equation \eqref{eq:feq} is the same as the hydrodynamic velocity obtained in equation \eqref{eq:LB_uhdydro}. 
This isn't the case for other forcing schemes.

The addition of a heat source term (thermal forcing term) is performed according to \citep{Thermal_forcing_Seta} with 
\begin{equation}
    q_i =   \Bigl( 1 - \frac{1}{2 \tau_g}\Bigr) w_i  \Phi \Delta t,
\end{equation}
where $\Phi=-\alpha(T-T_p)$ is the required source term. 
The temperature is then obtained at each lattice grid point from the thermal populations $g_i$ as 
\begin{equation}
    T = \sum_i g_i + \Bigl( 1 - \frac{1}{2 \tau_g}\Bigr)  \Phi.
\end{equation}

The no-slip boundary condition for the velocity at the top and bottom walls are imposed using the bounce-back method \citep{ladd1994numerical}. 
The adiabatic boundary condition for the top and bottom walls are imposed using the Inamuro method for setting the normal flux at a boundary for an advected scalar in a fluid \citep{yoshino2003lattice} by setting the flux equal to $0$. 
\section{Effects of varying \texorpdfstring{$\alpha_0$}{Lg} and \texorpdfstring{$c$}{Lg}}\label{sec:AppB_AlphVar}
It is clear from the main text that an increase in the number of particles $N_p$ strongly pushes the system towards the convective configuration while increasing $T_+$ weakly causes the system to tend towards stability. 
The other ways a phase change from a stable configuration to a convective configuration can be triggered is by increasing the fluid-particle coupling strength $\alpha_0$ or the size of the particle $c$, both of which serve to increase the typical velocity $u_0$. 

\begin{figure}
    \centering
    \begin{subfigure}{0.49\columnwidth}
        \includegraphics[width=\columnwidth]{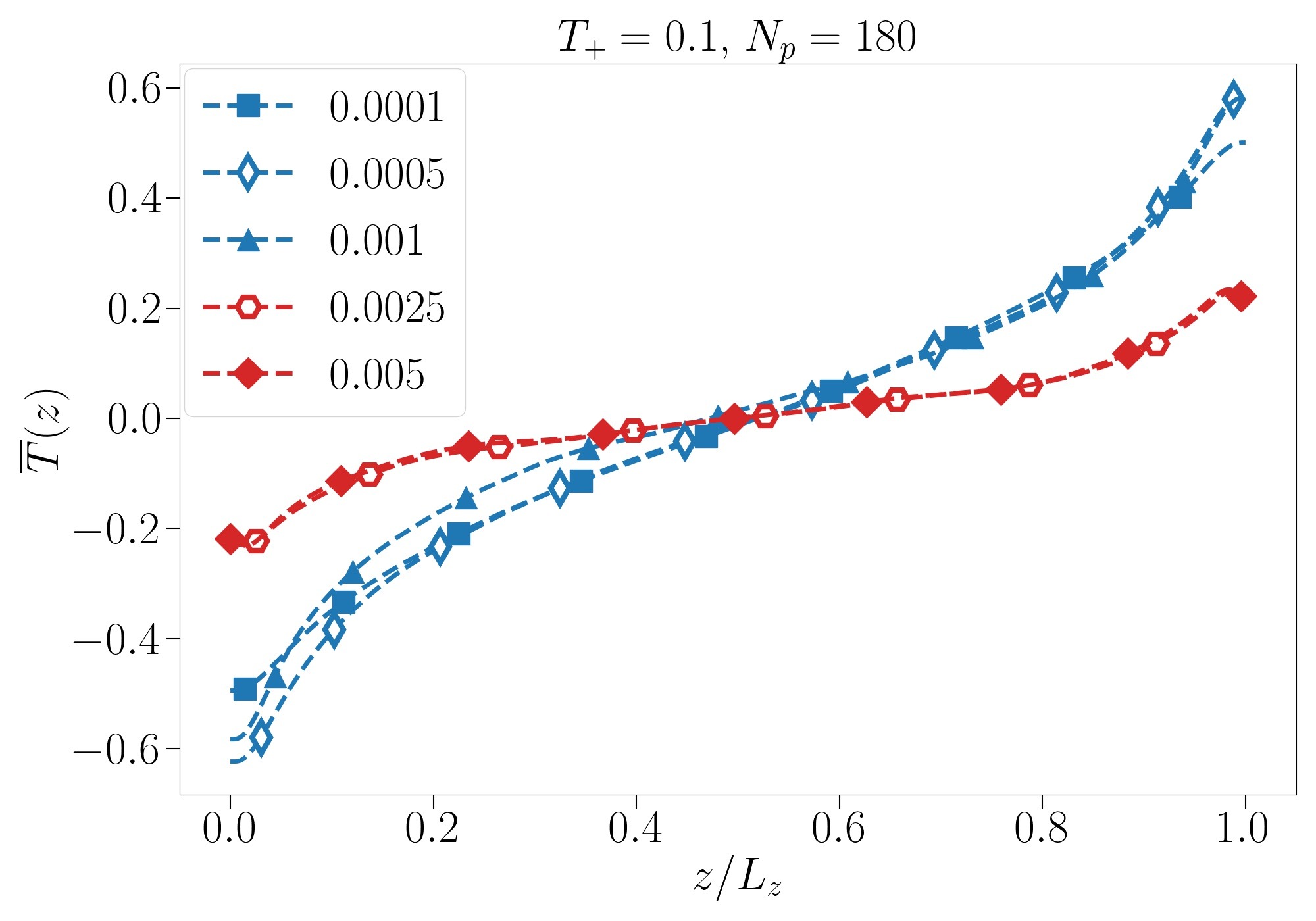}
        \caption{}
    \end{subfigure}
    \begin{subfigure}{0.49\columnwidth}
        \includegraphics[width=\columnwidth]{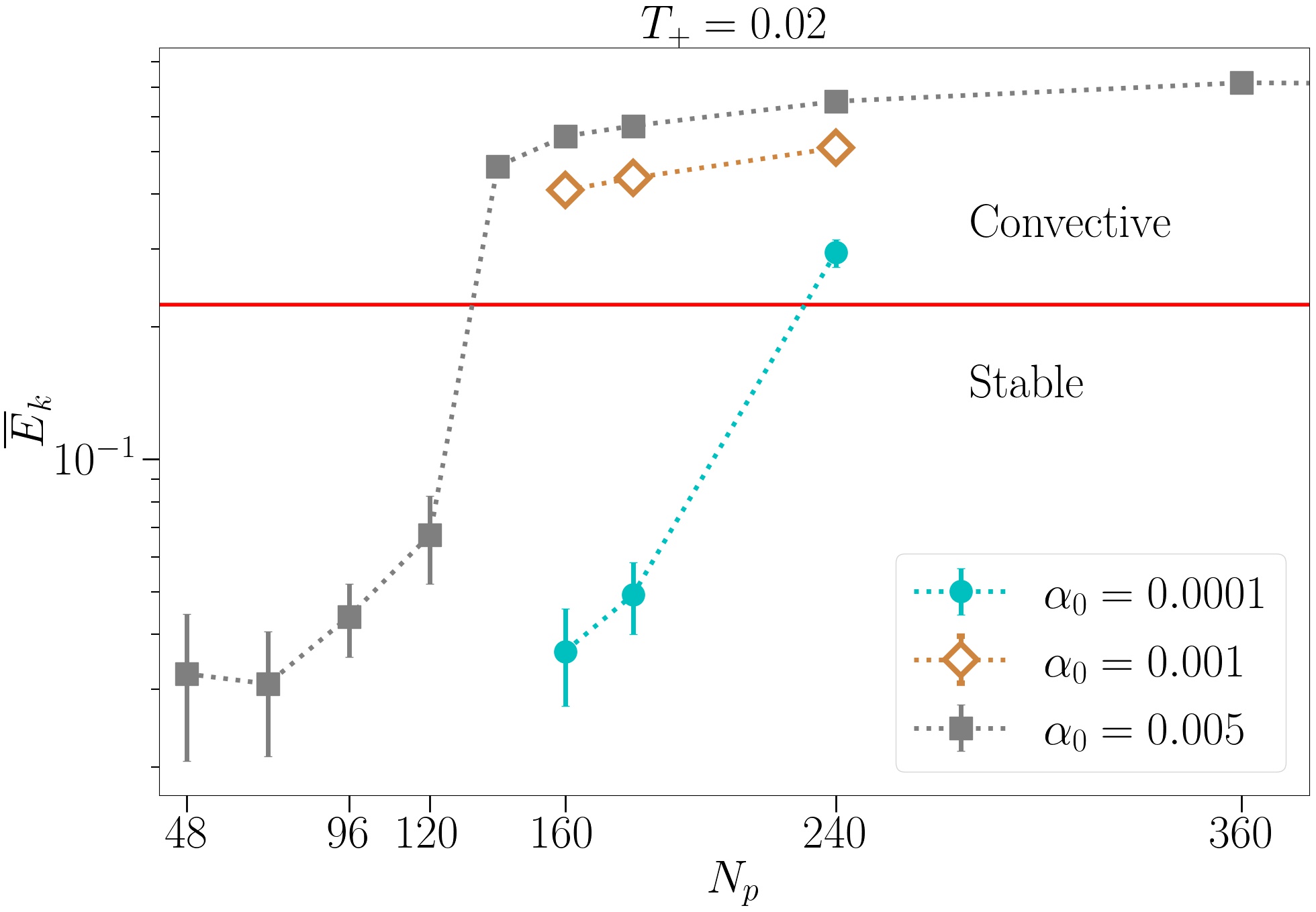}
        \caption{}
    \end{subfigure}
    \caption{(a) Normalised vertical temperature profiles for $T_+ = 0.01$, $N_p = 180$ for different $\alpha$. The red curves correspond to convective flows while the blue curves represent the stable flows. (b) Normalised TKE for $T_+ = 0.02$ plotted against $N_p$ for 3 values of $\alpha_0$. Horizontal red line represents $E^0_k = 0.225$. }
    \label{fig:alph_var}
\end{figure}

\begin{figure}
    \centering
    \includegraphics[width = 0.6\linewidth, keepaspectratio]{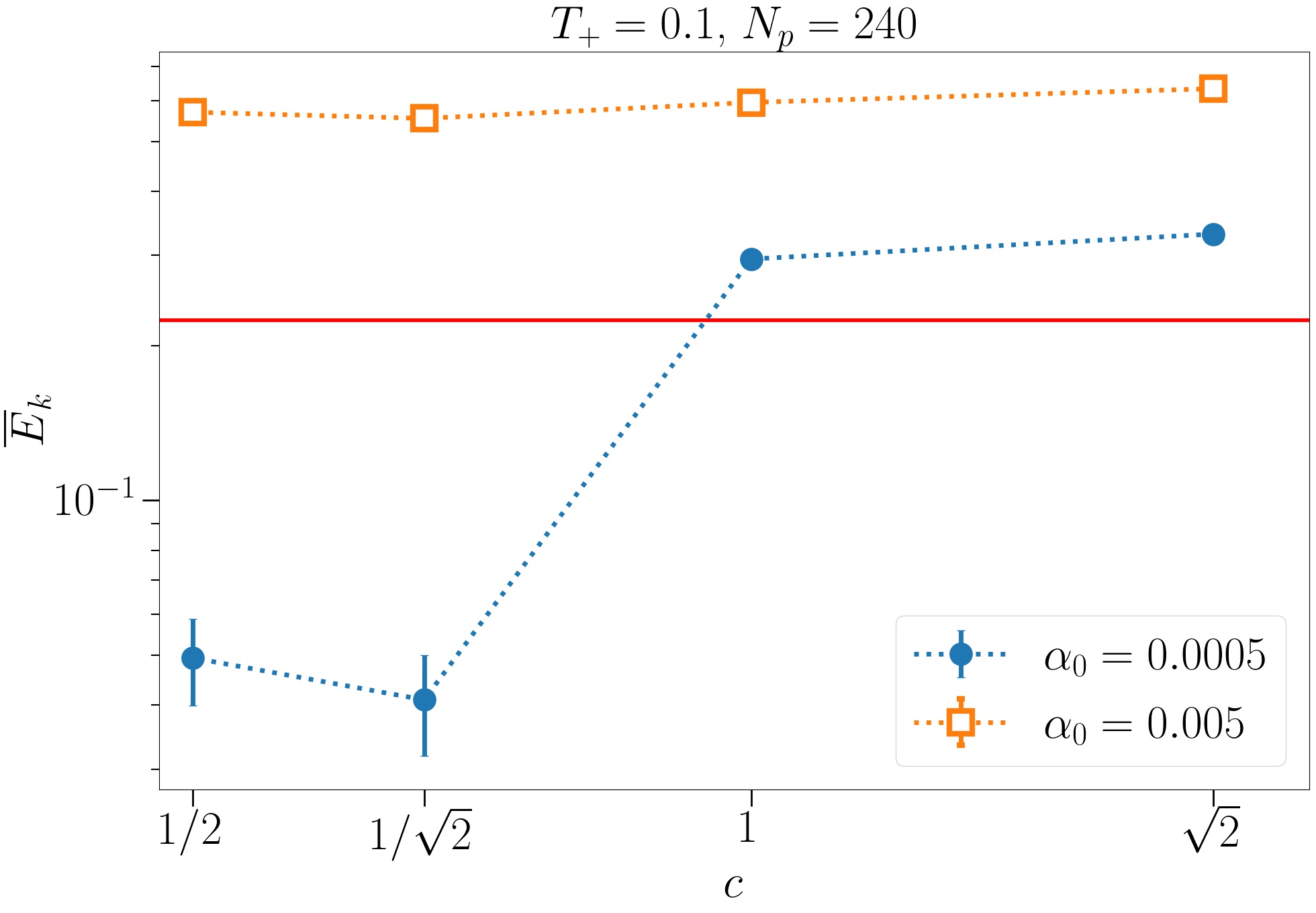}
    \caption{Normalised TKE for varying virtual particle size $c$ for two different $\alpha_0$.}
    \label{fig:c_var}
\end{figure}

The former effect can be gauged in figure~\ref{fig:alph_var}. 
In panel (a), we see the behavior of the temperature profile for varying $\alpha_0$. 
It has already been seen that the stable regime is characterised by a strongly stable temperature profile while the convective regime is characterised by a weakly stable temperature gradient. 
The temperature profile remains nearly identical for changing values of $\alpha_0$ except when the flow changes from stable (blue curves) to convective (red curves). 
We also note that the time taken for the flow to relax from the initial unstable configuration (see equation~\eqref{eq:initTprof}) to the eventual stationary state is larger for smaller $\alpha_0$. 
It indicates that for a given temperature scale $T_+$ and $N_p$, there exists a temperature difference $\overline{\Delta T}$ for which the flow is stable independent of $\alpha_0$. 
Panel (b) of the same figure where we plot the average normalised TKE $\overline{E}_k$ shows the transition from stable to convective for 3 different $\alpha_0$. 
That the increase in TKE corresponds to the transition from stable to convective was verified from visualisations of the flow field as well as the strength of the large-scale circulation as already discussed in Section~\ref{subsec:Lsc-Heat}. 
We see that decreasing $\alpha_0$ increases the $N_p$ of the transition and still note that the empirical value of $E_k^0$ for the transition holds. 

Increasing $c$ too shows a similar effect, as clear in figure~\ref{fig:c_var} where keeping the other parameters fixed, a transition to convective configuration is triggered by enlarging the size of the individual virtual particle. 

\section{Thermal Dissipation}\label{sec:AppC_ThermD}
We define the thermal dissipation rate as standard in the turbulence literature as
\begin{equation}
    \epsilon_T \equiv \kappa \bigl\langle(\partial_i T(\boldsymbol{x},t))^2\bigr\rangle_V,
\end{equation}
and note that in the statistically stationary regime, the thermal dissipation is equal to the thermal injection. 
We have the heat equation given by 
\begin{equation}
  \partial_t T + \uu \cdot \nab T = \kappa \nabla^2 T + \Phi.
  \label{eq:Nudged-heat}
\end{equation}
Following \citep{siggia1994high} and as shown explicitly by \citet[pp. 5-7]{ching2014statistics} for the Rayleigh-B\'enard convection, we multiply equation \eqref{eq:Nudged-heat} with $T$ and average over the entire domain and time to give
\begin{multline}
    \frac{1}{2}\frac{d \langle T^2 \rangle_{V,t}}{dt} + \frac{1}{2} \bigl\langle \uu \cdot \nab (T^2) \bigr\rangle_{V,t} - \bigl\langle \Phi T \bigr\rangle_{V,t} \\
    = \kappa \bigl\langle T \nabla^2 T \bigr\rangle_{V,t} = \kappa \bigl\langle \nab \cdot (T \nab T) \bigr\rangle_{V,t} - \kappa \bigl\langle | \nab T |^2 \bigr\rangle_{V,t},
    \label{eq:Tdot_heat}
\end{multline}
and then use the stationary condition ($\partial_t \langle \cdot \rangle_{V,t} = 0$) and the incompressibility ($\nab \cdot \uu = 0$) condition to give 
\begin{equation}
    \bigl\langle \uu \cdot \nab (T^2) \bigr\rangle_V 
    = \bigl\langle \nab \cdot (\uu T^2)\bigr\rangle_V = 0.
\end{equation}
Then, equation \eqref{eq:Tdot_heat} becomes 
\begin{equation}
    \kappa \bigl\langle | \nab T |^2 \bigr\rangle_{V,t} = \kappa \bigl\langle \nab \cdot (T \nab T) \bigr\rangle_{V,t} 
    + \bigl\langle \Phi T \bigr\rangle_{V,t},
\end{equation}
or 
\begin{equation}
    \epsilon_T = \kappa \bigl\langle \nab \cdot (T \nab T) \bigr\rangle_{V,t} 
    + \bigl\langle \Phi T \bigr\rangle_{V,t}
\end{equation}
The first term of $\epsilon_T$ can further be simplified using the Gauss theorem and writing it in terms of a surface integral
\begin{equation}
    \kappa \bigl\langle\nab \cdot (T \nab T) \bigr\rangle_{V,t} 
    = \frac{\kappa}{L_z} \biggl[ \Bigl\langle T \partial_z T \Bigr\rangle_{z=L_z} 
    - \Bigl\langle T \partial_z T \Bigr\rangle_{z=0}\biggr].
\end{equation}
In this study, we set $\partial_z T = 0$ at $z=0$ and $z=L_z$. Thus, finally we get simply
\begin{equation}
    \epsilon_T = \langle \Phi T \rangle_{V,t}.
\end{equation}

\bibliographystyle{jfm}
\bibliography{jfm}

\end{document}